\documentclass{JHEP3}
\usepackage{latexsym}

\def\AdSs5{$AdS_5$}
\def\AdS5s5{$AdS_5 \times S^5$}

\def\RR{{$RR$}}

\def\calB{{\cal B}}

\def\calP{{\cal P}}

\def\calN{{\cal N}}

\newcommand{\ie}{{\it i.e.~}}

\newcommand{\be}{\begin{equation}}
\newcommand{\ee}{\end{equation}}
\newcommand{\ba}{\begin{eqnarray}}
\newcommand{\ea}{\end{eqnarray}}

\newbox\SlashedBox
\def\fs#1{\setbox\SlashedBox=\hbox{#1}
\hbox to
0pt{\hbox to 1\wd\SlashedBox{\hfil/\hfil}\hss}{#1}}
\def\hboxtosizeof#1#2{\setbox\SlashedBox=\hbox{#1}
\hbox to
1\wd\SlashedBox{#2}}

\def\ms#1{\setbox\SlashedBox=\hbox{$#1$}
\hbox to 0pt{\hbox to
1\wd\SlashedBox{\hfil/\hfil}\hss}#1}

\def\t2{\tau_2}
\def\AdSS5{$AdS_5$}
\def\AdS5s5{$AdS_5
\times S^5$}

\def\calC{{\cal C}}

\def\calP{{\cal P}}

\def\IZ{\relax\ifmmode\mathchoice {\hbox{\cmss Z\kern-.4em
Z}}{\hbox{\cmss Z\kern-.4em Z}} {\lower.9pt\hbox{\cmsss Z\kern-.4em
Z}} {\lower1.2pt\hbox{\cmsss Z\kern-.4em Z}}\else{\cmss Z\kern-.4em
Z}\fi}

\def\tr{{\rm tr}}

\def\RR{{$RR$}}

\def\calA{{\cal A}}

\def\c1{{\chi^1}}

\def\N4{{\cal N}=4}
\def\half{{1\over 2}}
\def\nn{\nonumber}

\def\tr{{\rm tr}}

\def\x{{\bf x}}

\def\A{{\cal A}}

\def\ie{{\it i.e.}}
\def\half{{{1 \over 2}}}

\def\hn{{\hat n}}
\def\hp{{\hat p}}
\def\ttt{{\tilde t}}
\def\ts{{\tilde s}}
\def\tq{{\tilde q}}
\def\hm{{\widehat m}}
\def\hg{{\widehat g}}
\def\hom{{\widehat\omega}}
\def\pmb#1{\setbox0=\hbox{#1}%
 \kern-.025em\copy0\kern-\wd0
 \kern.05em\copy0\kern-\wd0
 \kern-.025em\raise.0433em\box0 }
\font\cmss=cmss10
\font\cmsss=cmss10 at 7pt
\def\rlx{\relax\leavevmode}
\def\Zop{\rlx\leavevmode\ifmmode\mathchoice{\hbox{\cmss Z\kern-.4em Z}}
 {\hbox{\cmss Z\kern-.4em Z}}{\lower.9pt\hbox{\cmsss Z\kern-.36em Z}}
 {\lower1.2pt\hbox{\cmsss Z\kern-.36em Z}}\else{\cmss Z\kern-.4em
 Z}\fi}

\def\bbbone {{\mathchoice {\rm 1\mskip-4mu l} {\rm 1\mskip-4mu l}
{\rm 1\mskip-4.5mu l} {\rm 1\mskip-5mu l}}}

\def\pmb#1{\setbox0=\hbox{#1}%
 \kern-.025em\copy0\kern-\wd0
 \kern.05em\copy0\kern-\wd0
 \kern-.025em\raise.0433em\box0 }
\font\cmss=cmss10
\font\cmsss=cmss10 at 7pt
\def\rlx{\relax\leavevmode}
\def\Cop{\relax\,\hbox{$\kern-.3em{\rm C}$}}
\def\Rop{\relax{\rm I\kern-.18em R}}
\def\Nop{\relax{\rm I\kern-.18em N}}
\def\Pop{\relax{\rm I\kern-.18em P}}

\def\Zop{\rlx\leavevmode\ifmmode\mathchoice{\hbox{\cmss Z\kern-.4em Z}}
 {\hbox{\cmss Z\kern-.4em Z}}{\lower.9pt\hbox{\cmsss Z\kern-.36em Z}}
 {\lower1.2pt\hbox{\cmsss Z\kern-.36em Z}}\else{\cmss Z\kern-.4em
 Z}\fi}

\def\m{{m}}

\def\xxx#1 {{hep-th/#1}}

\def\npb#1(#2)#3 { Nucl. Phys. {\bf B#1} (#2) #3 }
\def\rep#1(#2)#3 { Phys. Rept.{\bf #1} (#2) #3 }
\def\plb#1(#2)#3{Phys. Lett. {\bf #1B} (#2) #3}
\def\prl#1(#2)#3{Phys. Rev. Lett.{\bf #1} (#2) #3}
\def\physrev#1(#2)#3{Phys. Rev. {\bf D#1} (#2) #3}
\def\ap#1(#2)#3{Ann. Phys. {\bf #1} (#2) #3}
\def\rmp#1(#2)#3{Rev. Mod. Phys. {\bf #1} (#2) #3}
\def\cmp#1(#2)#3{Comm. Math. Phys. {\bf #1} (#2) #3}
\def\mpl#1(#2)#3{Mod. Phys. Lett. {\bf #1} (#2) #3}
\def\ijmp#1(#2)#3{Int. J. Mod. Phys. {\bf A#1} (#2) #3}

\title{The $D$-instanton and other supersymmetric $D$-branes in
IIB plane-wave string theory}
\author{Matthias R. Gaberdiel\footnote{On leave of absence from:
Department of Mathematics, King's College London, Strand,
London WC2R 2LS, UK.} \\
School of Natural Sciences, Institute for Advanced Study \\
Princeton, NJ 08540, USA \\
E-mail: \email{mrg@sns.ias.edu}, \email{mrg@mth.kcl.ac.uk}}
\author{Michael B. Green \\
Department of Applied Mathematics and Theoretical Physics \\
Wilberforce Road, Cambridge CB3 0WA, UK \\
E-mail:
\email{M.B.Green@damtp.cam.ac.uk}}

\abstract{A class of $D$-branes for the type IIB plane-wave background
is considered that preserve half the dynamical supersymmetries of the
light-cone gauge.  The $D$-branes of this type are the euclidean (or
instantonic) $(0,0)$, $(0,4)$ and $(4,0)$ branes (where $(r,s)$
denotes a brane oriented with $r$ axes in the first four directions
transverse to the $+,-$ light-cone and $s$ axes in the second four
directions). Corresponding lorentzian $D$-branes are $(+,-;0,0)$,
$(+,-;0,4)$ and $(+,-;4,0)$. These are constructed in two ways.  The
first uses a boundary state formalism which implements appropriate
fermionic gluing conditions and the second is based on a direct
quantisation of the open strings ending on the branes.   In
distinction to the $D$-branes considered earlier these
have massless world-volume fermions but do not possess
kinematical supersymmetries. Cylinder diagrams describing the overlap
between a pair of boundary states displaced by some distance are
evaluated. The open-string description of this system involves mode
frequencies that are, in general, given by irrational solutions to
transcendental equations. The closed-string and open-string
descriptions are shown to be equivalent by a nontrivial implementation
of the  $S$ modular transformation. A classical description of the
$D$-instanton (the $(0,0)$ case) in light-cone gauge is also given.}

\preprint{ KCL-MTH-02-25 \\ DAMTP-2002-139 \\ hep-th/0211122}

\begin{document}

\section{Introduction and review}
\label{intro}

The study of type IIB superstring theory in $pp$-wave  backgrounds
is a fertile area for investigating the properties of string theory with
nontrivial Ramond--Ramond (\RR) condensates.  In the simplest case,
the plane-wave background  with a constant flux of the Ramond--Ramond
(\RR) five-form field strength, there are 32 supersymmetries, which is
the maximal number. Perturbative superstring theory in this background
was formulated in \cite{met,mt}.
The metric has a $SO(4) \times SO(4)$ isometry
group which distinguishes the directions $x^I$ with $I=i = 1,2,3,4$
from the directions with $I= i'+4 =  5,6,7,8$.  In this notation the
light-cone directions are $x^\pm = (x^0 \pm x^9)/\sqrt{2}$, where
$x^0$ is time-like.  One of the main interests in studying such
theories is the connection, via the Penrose limit \cite{hulletal},
with $\calN=4$ supersymmetric Yang--Mills theory \cite{maldetal}.
A variety of generalizations of the maximally supersymmetric
plane-wave with less supersymmetry have been constructed.
Particularly interesting is the class of theories that are based on
$(2,2)$ world-sheet supersymmetry \cite{mm}.  These backgrounds have been
shown to be exact solutions of superstring theory to all orders in
$\alpha'$ \cite{bm}.  In these cases the
generic background, which is expressed in terms of a superpotential,
has less space-time supersymmetry and the string theory is governed by
an integrable two-dimensional system.

This paper will continue the study of $D$-branes which preserve
half of the dynamical supersymmetry of the maximally supersymmetric
plane-wave background. For much of the time we will be discussing
instantonic $D$-branes, which are defined by euclidean embeddings of
$(p+1)$-dimensional world-volumes.  These are the cases in which the
light-cone directions $x^\pm$ are orthogonal to the brane. We will
adopt a notation (similar to that in \cite{st}, which provides a useful
overview of various $D$-branes that arise in this background) in which these
instantonic branes are denoted $(r,s)$-branes ($r+s = p+1$), where $r$
and $s$ are the numbers of directions associated with the two $SO(4)$
factors in the transverse space. Our considerations apply equally well
to $Dp$-branes with lorentzian signature which will be denoted
$(+,-;r,s)$-branes  $(r+s= p-1)$.   Recently, it has been found that
there are also various `oblique' branes which are constrained to be
oriented in directions that couple the two $SO(4)$'s \cite{hk}. Such
branes, which cannot be classified as $(r,s)$-branes,  possess less
supersymmetry than those considered here. They arise naturally in the
backgrounds of \cite{mm}.

In section~\ref{boundstate} we will present the construction of the
branes in a light-cone formalism that extends the discussions of
\cite{bgg}.  From the closed-string perspective this generalizes
the flat-space construction in \cite{gg}, and starts with the
ansatz that the boundary state describing the brane is
annihilated by half the `dynamical' light-cone
supercharges\footnote{Conventions and notation are explained in
appendix~\ref{modprops}.},
\be
\left( Q_{\dot a} + i \eta\, M_{\dot a \dot b}\,
{\tilde Q}_{\dot b}\ \right)
|\!| (r,s),\eta\,\rangle\!\rangle = 0\,,
\label{dynamicalnn}
\ee
where the value of $\eta =\pm 1$ distinguishes a brane from an
anti-brane.  We will also assume that the bosonic coordinates satisfy
the standard boundary conditions in Dirichlet directions (but not
necessarily in Neumann directions since these may be affected by
the background \RR\ flux). As we will see these conditions also imply
that the boundary state preserves half the `kinematical' light-cone
supercharges,
\be\label{kinematicaln}
\left( Q_a + i \eta\, M_{ab}\, {\tilde Q}_b \right)
|\!| (r,s),\eta\,\rangle\!\rangle = 0\,.
\ee
This means that the fermionic modes satisfy the gluing conditions
\be
\left( S^a_0+ i \eta \, M_{ab} \, \tilde{S}^b_0 \right)
|\!| (r,s),\eta\,\rangle\!\rangle = 0\,, \qquad k\in\Zop \,.
\label{boundarycondii}
\ee
Importantly, we will {\it not} assume that the same condition
holds for the non-zero modes of the supercharge density, which was
the case in \cite{bp} and \cite{bgg}.

As in the flat-space case the matrix $M$ is  given by
\be
M_{ab} = \left( \prod_{I\in{\cal N}} \gamma^I \right)_{ab}\, ,
\qquad
M_{\dot{a}\dot{b}} =
\left( \prod_{I\in{\cal N}} \gamma^I \right)_{\dot{a}\dot{b}}\,,
\label{Mdot}
\ee
in the two inequivalent $SO(8)$ spinor representations. Here
${\cal N}$ denotes the set of directions for which the brane satisfies
a `Neumann' boundary condition. We will see that in the $pp$-wave
background there are two classes of maximally supersymmetric
$Dp$-branes, depending on the choice of $M$:\hfil\break
$\bullet{}$ {\bf Class I}. The first class is the one that was studied
in \cite{bp,dp,bgg} and arises when the matrix $M_{ab}$ satisfies
\be
\Pi M \Pi M = -1\, ,
\label{classone}
\ee
where $\Pi =\gamma^1\gamma^2\gamma^3\gamma^4$.  This
condition was considered in the cases of instantonic $D$-branes
in \cite{bp} and lorentzian $D$-branes in \cite{dp}.  The branes of
this kind are of the form $(r,r+2)$ and $(r+2,r)$.  They preserve half
of the dynamical supersymmetries as well as half of the kinematical
supersymmetries. There are two types of $M$ that satisfy
(\ref{classone}),\hfil\break
\be
(i) \qquad \qquad \Pi M = - M\Pi, \qquad M^T = M\, ,
\label{acomm}
\ee
which includes the cases $(3,1)$, $(1,3)$, or
\be
(ii) \qquad\qquad \Pi M = M \Pi, \qquad M^T = -M\,,
\label{anticom}
\ee
which includes the cases $(2,0)$, $(0,2)$, $(4,2)$, $(2,4)$.
In all of these cases the open-string sector preserves eight
components (\ie\ half) of both the dynamical and kinematical
supersymmetries.  A characteristic feature of this class is that the
kinematical conditions (\ref{kinematicaln}) are not preserved as a
function of $x^+$ since the commutator with the light-cone hamiltonian
has the form
\be
[H,Q_a + i \eta\, M_{ab}\,{\tilde Q}_b] =
{m\, \eta \over 2 p^+} (\Pi M^t)_{ab}
\left(Q_b  - i \eta\, M_{bc}\,{\tilde Q}_c \right) \,.
\ee
In this case the open-string theory has a mass term in its hamiltonian
of the form $S_0 M\Pi S_0$ \cite{dp}, and the ground state is an
unmatched boson.

\noindent $\bullet{}$ {\bf Class II}.
The second class arises when  the matrix $M_{ab}$
satisfies
\be
\Pi M \Pi M = 1\, ,
\label{classtwo}
\ee
a possibility that was not considered in \cite{bp,dp,bgg} but
arose in the supergravity analyses of \cite{st,bmz}.  Those
branes in this class which preserve half of the dynamical
supersymmetries possess no open-string kinematical supersymmetries.
There are two types of $M$ satisfying
(\ref{classtwo}),\hfil\break
\be
(i) \qquad \qquad \Pi M = M\Pi, \qquad M^T = M\, ,
\label{acomn}
\ee
which includes the cases $(0,0)$, $(2,2)$, $(4,0)$, $(0,4)$, $(4,4)$ or
\be
(ii) \qquad\qquad \Pi M = - M \Pi, \qquad M^T = -M\,.
\label{anticon}
\ee
which includes the cases $(1,1)$, $(3,3)$.  In these cases
\be
[H,Q_a + i \eta\, M_{ab}\,{\tilde Q}_b]
= - {m \, \eta \over 2 p^+} (\Pi M^t)_{ab}
\left( Q_b + i \eta\, M_{bc}\, {\tilde Q}_c \right)\,,
\ee
and thus the kinematical conditions (\ref{kinematicaln}) {\it are}
preserved as a function of $x^+$.  In this case
$S_0 M\Pi S_0\equiv 0$ and the open-string  mass term vanishes. The
ground states then form a degenerate supermultiplet.

In section~\ref{boundstate} we will obtain the boundary states for the
$(0,0)$-brane (or $D$-instanton), which is a class II brane.  We will
also  evaluate the overlap of the $(0,0)$ with itself, with the
$\overline{(0,0)}$ (the anti $D$-instanton),
and with $(r,r+2)$-branes. In this manner we
determine the cylinder diagrams that describe the interaction
energies (more accurately, the `interaction actions') of these
instantonic objects.

The relevant cylinders have parameter length  $X^+$, which is the
difference between the $x^+$ positions of the two instantonic
$D$-branes, and circumference  $2\pi p^+$. The relevant modular
parameter of the cylinder is therefore $X^+/2\pi p^+$. The light-cone
gauge in the closed-string channel is chosen such the $x^+$ is
proportional to the world-sheet time coordinate $\tau$ and thus $\tau$
parametrises the direction along the finite length of the cylinder,
while the world-sheet space coordinate $\sigma$ is the periodic
variable along the circular direction. In order to understand the
relation to the open-string description it is necessary to perform
a Wick rotation, replacing $\tau$ by
$i\tau$ as well as $x^+$ by $ix^+$, so that the world-sheet theory, as
well as the space-time theory, become euclidean. (While it is not
clear how to perform the Penrose limit for a euclidean theory, it is
certainly possibly to Wick rotate the resulting plane wave background.)
The two world-sheet coordinates $\sigma$ and $\tau$ then appear on an
equal footing, and the cylinder diagram can be written as a
function of the modular parameter (see \cite{bgg} for more details)
\be
t = {X^+ \over 2\pi p^+} \,,
\ee
where $X^+$ is the $x^+$ separation between the two instantonic
$D$-branes after the Wick rotation.\footnote{That is,      ,
$X^+=i X^+_0$, where $X^+_0$ is the original distance between the
$D$-instantons that is analytically continued to purely imaginary
values in performing the Wick rotation.}  More specifically, for the
class I branes discussed in \cite{bgg}, the cylinder diagrams can be
expressed in terms of ratios of powers of functions $f^{(m)}_i(t)$
($i=1,2,3,4$) whose definition is given in appendix~\ref{appa}.

In order to identify the cylinder diagram with an open
string one-loop diagram,  the world-sheet
parameters of the open string, $\tilde\tau$ and $\tilde\sigma$, are
then  identified with
those of the closed string so that $\tilde\tau=\sigma$ and
$\tilde\sigma=\tau$. This corresponds to choosing the light-cone gauge
for the open string where  $x^+$ is proportional to $\tilde\sigma$. In
the euclidean description where $\tilde\tau$ and $\tilde\sigma$ appear
on the same footing, this is a consistent gauge choice.

In this open-string light-cone gauge the total length of the string in
the $\tilde \sigma$ direction is $X^+$, while the proper time around
the loop is $2\pi p^+$. This  means that $X^+$ and  $2\pi p^+$ are
interchanged relative to the usual formulae that would hold for an
open string ending on a time-like (lorentzian signature) brane. In
particular, while the `mass parameter' of the usual open string
is $m=2\pi \mu p^+$, the `mass parameter' in the light-cone gauge
appropriate to the instantonic branes is $\hm = \mu X^+$.
Furthermore, the modular parameter in the open string description is
\be
\ttt = {1\over t} = {2\pi p^+ \over X^+} \,.
\ee
The requirement that the cylinder diagram can equally well be
described in terms of the open and closed string point of view implies
that the amplitudes must transform appropriately under the $S$ modular
transformation $t\mapsto \ttt=1/t$. For the case of the class I branes
this turned out to be the case \cite{bgg} because the functions
$f_i^{(m)}$ transform as
\be
f^{(m)}_1(t) = f^{(\hm)}_1(\tilde t)\,, \qquad
f^{(m)}_2(t) = f^{(\hm)}_4 (\tilde t)\,,\qquad
f^{(m)}_3(t) = f^{(\hm)}_3(\tilde t)\,,
\label{ftrans}
\ee
where $\hm = m t = \mu X^+$ is the mass parameter in the open string
description, as discussed above. In the limit $m\to 0$ these
functions become the standard $f_i$ functions (except for a subtlety
involving the zero modes for $f_1$) that are used to describe the
cylinder diagrams of the flat space theory.

As usual, the discussion of section~\ref{boundstate} does not fix the
overall normalisation of the boundary states. This normalisation
is left to be determined by relating the closed string calculation to
the canonically normalised open string calculation as discussed
above. The construction of the open string in section~\ref{openstring}
is somewhat subtle compared to previous cases because the open string
fermions have states with frequencies $\hom_n  = \sqrt{n^2 + \hm^2}$,
where $n$ is generically  an irrational number that is a solution of
either
\be
{n + i\hm \over n - i\hm} + e^{2\pi i n} =0\,,
\label{calpplusdef}
\ee
or
\be
{n -  i\hm \over n + i\hm} + e^{2\pi i n} =0\,,
\label{calpmindef}
\ee
and $\hm=\mu X^+$ is the mass-parameter in the open string light cone
gauge. The space of solutions to the first of these equations will be
called $\calP_+$ while the second equation defines the space
$\calP_-$. Although these equations also include the value $n=0$, this
is not an allowed value for the open-string fermionic oscillators (as
will be explained later).

In section~\ref{oneloop} and appendix~\ref{modprops} we will show that
the open-string expression coincides with that derived from the
boundary state approach, \ie\ that the cylinder diagram transforms
consistently under the $S$ modular transformation condition as in
\cite{bgg}. The fermionic contributions to the cylinder diagrams
require further functions $g^{(m)}_2$ and $\hg^{(\hm)}_4$.  In
appendix~\ref{appa} it is explained that these functions transform
into each other under the $S$ modular transformation,
\be
g^{(m)}_2(t) = \hg^{(\hm)}_4(\tilde t)\,.
\label{gtrans}
\ee
The proof is considerably more subtle than that for the $f^{(m)}_i$
functions  because the frequencies of the open string excitations in
this case take the irrational values discussed above.

A similar analysis can also be performed for the lorentzian branes, as
discussed in \cite{bgg}. In this case the open-string light-cone gauge
is the usual one in which $x^+ = 2\pi \tilde \tau p^+$ and, the
parameter length of the string is $2\pi p^+$.  Following a Wick
rotation, the closed-string description is one in which  the length of
the string in the $\sigma$-direction is now $X^+$, while the proper
time interval is $2\pi p^+$. Thus the `mass parameter' in the open
string is the usual $m$, while now the `mass parameter' in
the closed string is $\hm$. At any rate, the actual calculations are
virtually identical to the corresponding calculations in the euclidean
case, and we shall therefore not discuss them in detail.

In section~\ref{zerfour}  we will consider the cases of the
$(0,4)$ and $(4,0)$ branes, starting from the conditions on the
boundary states.  In this case there is no consistent set of  gluing
conditions unless the bosons satisfy modified Neumann conditions in
directions tangent to the brane.
This is, of course, expected since these branes couple to the anti
self-dual four-form background potential associated with the
constant \RR\ five-form field strength.  The $(0,4)$ and $(4,0)$
branes do not have supersymmetric anti-branes -- the obvious
candidates turn out to be branes for the theory in which the
\RR\ four-form potential is self-dual instead of anti self-dual.
The overlap between a $(4,0)$
and a separated $(2,0)$ is obtained and used to derive the appropriate
boundary conditions on the open strings.  The modified Neumann conditions
translate into those expected on the basis of the Born--Infeld
action and given in \cite{st} and \cite{hk} (where they were obtained
by requiring $(2,2)$ world-sheet supersymmetry).

The $(0,0)$, $(4,0)$ and $(0,4)$ instantonic branes mentioned above
seem to be the only class II branes that lead to a conserved dynamical
supercharge in the open string. The same statements hold for the
$(+,-;0,0)$, $(+,-;4,0)$ and $(+,-;0,4)$ lorentzian
branes. Somewhat tantalizingly, there is another construction
of a $(0,4)$ and $(4,0)$ boundary state, as well as of a $(4,4)$
boundary state, that  satisfies (\ref{dynamicalnn}) and
(\ref{kinematicaln}) in the closed string sector. However, the corresponding
open strings do not possess a conserved dynamical supercharge, and
there are various indications that these boundary states do not
actually satisfy the open-closed duality relation. It would be
interesting to understand the meaning (if any) of these additional
coherent states. Finally, from our analysis it seems that
(\ref{dynamicalnn}) and (\ref{kinematicaln}) do not have solutions in
any of the other potential class II cases ($(1,1)$, $(2,2)$, and
$(3,3)$), which agrees with the supergravity analysis in
\cite{bmz}.

This conclusion is supported by the recent paper \cite{hk} which
analyzed $D$-branes in the generalized $pp$-wave backgrounds of
\cite{mm} that can be expressed in terms of $(2,2)$ world-sheet
supersymmetry. The branes considered there are ones which preserve
some fraction of this reduced supersymmetry. In
section~\ref{twotwosusy} we will argue that all of the potential
class II examples are of this type, and that their presence or absence
coincides with the results of \cite{hk}.  The additional `oblique'
branes of that paper preserve less than eight components of the
dynamical supersymmetry and should be seen by a small generalization
of our boundary state gluing conditions although we have not done that.

As in the flat ten-dimensional case, the $(0,0)$-brane couples to the
dilaton and the \RR\ pseudoscalar. In section~\ref{classinst} we will
review the supergravity arguments that imply the presence of the
$D$-instanton in flat space.  We will also give an interpretation of
the flat-space solution in the `euclidean' light-cone gauge, \ie\
after a Wick rotation in $x^+$. Making use of the conformal flatness
of the plane-wave background together with the vanishing of the scalar
curvature it is straightforward to see how the $(0,0)$-brane arises as
a solution in the plane wave case. However, it is less obvious how to
interpret the effect of this solution on the plane-wave dynamics.
Finally, in section~\ref{discuss} we will review these results and
make additional comments.

\section{The $(0,0)$  ($D$-instanton) boundary state and its overlaps}
\label{boundstate}

The  class I boundary states were constructed in \cite{bp} in a manner
that mimics the flat-space light-cone gauge construction \cite{gg} and
preserves a complex combination of the closed-string supercharges.  In
this approach the light-cone $x^\pm$ directions are taken to be
transverse to the brane world-volume which means that the description
is appropriate for euclidean branes with $(p+1)$-dimensional
world-volumes. The instantonic branes considered in \cite{bp} and
\cite{bgg} are of the form $(r,r+2)$ and $(r+2,r)$.  A similar set of
lorentzian signature $Dp$-branes of the form $(+,-;r,r+2)$ and $(+,-;r+2,r)$,
was obtained in \cite{dp} by analysis of the open-string sector.
These cases are those of
class I in the notation of section~\ref{intro}.  In other words, they are
ones for which the matrix $M$ satisfies $\Pi M\Pi M = -1$.
In this section we will consider the boundary state approach for
branes in class II where $M$ satisfies $\Pi M\Pi M = +1$.  In particular,
we will obtain a boundary state corresponding for the $(0,0)$ case  (the
$D$-instanton).  This will be generalized in section~\ref{zerfour}
to the $(4,0)$ and $(0,4)$ cases (euclidean $D3$-branes) which are
special because they couple to the constant background self-dual \RR\
flux.  These examples fit in well with the analysis of \cite{hk}
which is based on $(2,2)$ world-sheet supersymmetry, whereas the
class I branes are not invariant under this sub-symmetry (as
will be discussed further in section~\ref{twotwosusy}).

We are interested in constructing a  boundary state for a
$(r,s)$-brane located at a transverse position $x_{0}^I=y_t^I$.
The Dirichlet gluing conditions for the bosons take
the form
\ba
\left( \alpha^I_k - \tilde\alpha^I_{-k} \right)
|\!| (r,s), {\bf y}_t\,\rangle\!\rangle  & = & 0\,,
\qquad k\in\Zop\,, \nn\\
\left( \bar{a}^I_0 - a^I_0 + i \sqrt{2m} y_t^I \right)
|\!| (r,s), {\bf y}_t\,\rangle\!\rangle &\equiv &
- i\sqrt{2m}\, (x_{0}^I - y_t^I) \,
|\!| (r,s), {\bf y}_t\,\rangle\!\rangle = 0 \,,
\label{boundarycond}
\ea
where $I$ is a Dirichlet direction. Initially, the gluing conditions
for the Neumann directions are assumed to be
\ba
\left( \alpha^J_k + \tilde\alpha^J_{-k} \right)
|\!| (r,s), {\bf y}_t\,\rangle\!\rangle  & = & 0\,,
\qquad k\in\Zop\,, \nn\\
\left( \bar{a}^J_0 + a^J_0 \right)
|\!| (r,s), {\bf y}_t\,\rangle\!\rangle & = & 0\,.
\label{boundarycondN}
\ea
Furthermore, the boundary state must be annihilated by a certain
linear combination of the dynamical supercharges (defined in
appendix~\ref{notate})
\be
\left( Q_{\dot a} + i\,\eta\, M_{\dot a\dot b}\,
{\tilde Q}_{\dot b}\ \right)
|\!| (r,s),{\bf y}_t\,\rangle\!\rangle = 0\,,
\label{dynamical}
\ee
where $M_{\dot a\dot b}$ is given as in (\ref{Mdot}).
Here $\eta=\pm 1$ distinguishes the $(r,s)$-brane from the
$\overline{(r,s)}$ (the anti-brane). The gluing conditions for the
fermions are uniquely determined by (\ref{boundarycond}) and
(\ref{dynamical}). However, the resulting conditions do not yet
automatically imply that (\ref{dynamical}) actually holds. This
constraint imposes additional restrictions on the set of $(r,s)$
branes, and may require a modification of the Neumann conditions
(\ref{boundarycondN}), as will be described in section~\ref{zerfour}.

By considering the commutator of (\ref{dynamical})
with $x^I_0$ it follows that the boundary state must be annihilated by
\be
\left( S^a_0 + i\, \eta\, M_{ab}\,\tilde{S}^b_0 \right)
|\!| (r,s),{\bf y}_t\,\rangle\!\rangle = 0\,,
\label{kinematical}
\ee
which implies that a complex combination of the kinematical
supersymmetries is preserved by the boundary state.
With this gluing condition it follows that the Dirichlet zero mode
part of $Q_{\dot a} + i\,\eta\, (M \tilde Q)_{\dot a}$ (\ie\ the terms
proportional to $x_0^I$ and $p_0^I$ for which $I$ is a Dirichlet
direction) annihilates the boundary state for a class II brane
irrespective of the transverse position ${\bf y}_t$. On the other
hand, for class I branes, this condition is only satisfied if
$y_t^I=0$ (unless we modify the gluing conditions (\ref{boundarycond})
and  (\ref{boundarycondN})).

The gluing condition relating the non-zero modes $S^a_n$
and $\tilde{S}^b_{-n}$ at the boundary can be determined by first
writing the non-zero mode part of the dynamical supercharge in
(\ref{dynamical}) explicitly as
\begin{eqnarray}
&&
\sum_{n=1}^\infty
\left( c_n \gamma^{I}
       (\alpha_{-n}^I S_n  + \alpha_n^I S_{-n} )
+ {\eta\, m\over 2 \omega_n c_n }
 \left(M\gamma^I\Pi\right)
       (\alpha_{-n}^I S_n  - \alpha_n^I S_{-n})
\right) \label{q+i} \\
& & +\, i\,\left[
\sum_{n=1}^\infty
\left(
\eta c_n  M \gamma^{I}
       ({\tilde \alpha}_{-n}^I {\tilde S}_n
        +{\tilde \alpha}_n^I {\tilde S}_{-n} ) +
{m \over 2\omega_n c_n }
(\gamma^I\Pi)_{\dot a b}
       ({\tilde \alpha}_{-n}^I {\tilde S}_n^b
        -{\tilde \alpha}_n^I {\tilde S}^b_{-n} )
\right)\right] \, . \nn
\end{eqnarray}
Using the fact that the boundary state is annihilated by
$\alpha^I_{-n} -\tilde{\alpha}^I_n$ (for each Dirichlet direction $I$)
this equation can be solved to determine the modes $\tilde S_n$ in
terms of $S_n$. The solutions depend on the properties of
$M$. However, by construction, these equations guarantee so far
only that the Dirichlet part of the supercharges (\ie\ the terms that
are proportional to $\alpha^I_n$ or $\tilde{\alpha}^I_n$ with $I$ a
Dirichlet direction) annihilate (\ref{dynamical}); whether the full
equation (\ref{dynamical}) holds has to be analysed case by case.

{\bf Class I}.  In this case $\Pi M \Pi M = -1$, which is the
condition assumed in \cite{bp,bgg}. The solution of (\ref{dynamical})
is the same as in flat space,
\be
\label{fermions11}
\left(S^a_n + i \eta\, M_{ab}\,
 \tilde{S}^b_{-n} \right)
|\!| (r,r+2),{\bf y}_t=0\,\rangle\!\rangle = 0  \,,
\ee
which implies that the density of $(Q_a + i\eta (M \tilde Q)_a)$
annihilates the boundary state.  The cases with $(r+2,r)$ work in
an equivalent fashion. With these gluing conditions it is then easy to
see that the boundary state satisfies (\ref{dynamical}) with the
standard gluing conditions for the Neumann directions
(\ref{boundarycondN}).

{\bf Class II}.  When $\Pi M \Pi M = +1$ the gluing conditions
for the fermionic modes can again be determined as above, but now
(\ref{dynamical}) is only satisfied provided that
$(r,s)=(0,0)$.\footnote{Here we are assuming that the $D$-brane has at
least one transverse Dirichlet direction; it is also possible to solve
these equations for the case of the $(4,4)$-brane, but 
the resulting boundary state does not have a sensible flat 
space limit, and is probably inconsistent.}
Furthermore, if one
relaxes the conditions (\ref{boundarycondN})  there are also solutions
for the cases $(4,0)$ and $(0,4)$ which will be discussed in
section~\ref{zerfour}.

For the case of the $(0,0)$-brane, the matrix $M$ is simply
$M=\bbbone$, and the condition ({\ref{dynamical}) can be expressed in
the form
\be\label{fermions}
\left[\left(\bbbone + {\eta\,m\over 2 \omega_n c_n^2} \Pi \right)_{ab}
S^b_n
+ i\,\eta\,
\left(\bbbone - {\eta\, m\over 2 \omega_n c_n^2} \Pi \right)_{ab}
\tilde{S}^b_{-n} \right] |\!| (0,0),{\bf y}\,\rangle\!\rangle = 0
\ee
for all $n\in\Zop$, $n\ne 0$. Note that since $\omega_{-n}=-\omega_n$
and $c_{-n}=c_n$, this formula gives the correct expression both for
positive and negative $n$. This can be simplified using
\be\label{relation}
\left(\bbbone + {\eta\,m\over 2 \omega_n c_n^2} \Pi \right)_{ab}
\left(\bbbone - {\eta\,m\over 2 \omega_n c_n^2} \Pi \right)_{bc} =
{2 n \over m^2} (\omega_n -n ) \delta_{ac}\,,
\ee
which leads to the conditions
\be\label{fermions1}
\left(S^a_n + i \eta\, R_n^{ab}\,
 \tilde{S}^b_{-n} \right)
|\!| (0,0),{\bf y}\,\rangle\!\rangle = 0  \,,
\ee
where $R_n$ is the matrix
\be\label{rdef}
R_n = {1\over n} \left( \omega_n \bbbone - \eta\, m \Pi
\right) \,.
\ee
It is worth noting that $R_n$ is not orthogonal, but that
\be\label{orthogonal}
R_n R_{-n}^T = \bbbone \,.
\ee
This equation is required to make the above gluing
conditions self-consistent.

The expression  for $R_n$ can be simplified further by
decomposing the $SO(8)$ spinors  $S_n$ and $\tilde{S}_n$ into spinors
of definite $SO(4)$ chiralities by defining
\be\label{plusminus}
S^+_n = {1\over 2} (1+\Pi) S_n\,, \qquad
S^-_n = {1\over 2} (1-\Pi) S_n\,,
\ee
so that $\Pi S^\pm_n = \pm S^\pm _n$, and similarly for
$\tilde{S}_n$. Then (\ref{fermions1}) can be rewritten as
\be
\left(S^\pm_n + i \eta\, R_n^{\pm} \tilde{S}^\pm_{-n} \right)
|\!| (0,0),{\bf y}\rangle\!\rangle = 0 \,,
\ee
where
\be\label{rpmdef}
R_n^{\pm} = {\omega_n \mp \eta\, m \over n}
= \sqrt{{\omega_n \mp \eta\,m \over \omega_n \pm \eta\, m}}\,.
\ee
Given this expression, it is easy to write down the full
boundary state
\be
|\!| (0,0),{\bf y},\eta\,\rangle\!\rangle = {\cal N}_{(0,0)}
\exp \left( \sum_{k=1}^{\infty}
{1\over \omega_k} \alpha^I_{-k} \tilde\alpha^I_{-k}
- i \eta R_k^+ S^+_{-k} \tilde{S}^+_{-k}
- i \eta R_k^- S^-_{-k} \tilde{S}^-_{-k}\right) \,
|\!| (0,0)\,\rangle\!\rangle_0\,,
\label{boundone}
\ee
where ${\cal N}_{(0,0)}$ is a normalisation constant that will turn
out to equal
\be\label{norma}
{\cal N}_{(0,0)} = (4\,\pi\, m)^2\,,
\ee
and the ground state component is
\be
|\!| (0,0)\,\rangle\!\rangle_0
= \left( |I\rangle |I\rangle
+ i \eta |\dot{a}\rangle |\dot{a}\rangle \right)\,
e^{- {m {\bf y}^2 / 2}}\, e^{\half a_0^I a_0^I - i \sqrt{2m}\,
y^I a^I_0 } |0\rangle_b \,.
\label{boundzero}
\ee
Here the first bracket describes the `fermionic' part of the ground
state, while the second part describes its `bosonic'.
This ground state is just the linear sum of the
dilaton and \RR\ scalar that enters the flat-space boundary
$D$-instanton state.  This zero-mode part is an
eigenstate of the closed-string hamiltonian, in contrast to the
ground-state factors in the class I branes discussed in
\cite{bgg}.
When $m\rightarrow 0$,
the gluing condition (\ref{fermions1}) reduces to the usual flat-space
result since $R_n=\bbbone$ for $m=0$.

\subsection{Cylinder diagrams involving the $D$-instanton}
\label{cyldiag}

Given the explicit description for the $D$-instanton boundary state we
can now evaluate the sum over cylindrical world-sheets with one
boundary on a $(0,0)$-brane (or $D$-instanton) while the other boundary
describes in turn a $(0,0)$-brane, a $\overline{(0,0)}$-brane (or
anti $D$-instanton), or one of the class I branes discussed
in \cite{bp,bgg}. From the closed-string point of view that is
considered in this section, this involves the overlap of two boundary
states with non-zero $p^+$ that are separated in the $x^+$ direction
by $X^+$, as well as the $x^I$ directions. The diagrams can also be
identified with a trace over the states of the open  strings joining
the separated $D$-branes as will be seen in the next section.
 The proof that the
closed-string construction gives the same expression as the
open-string construction is a consequence of the modular property of
the functions $g_2$ and $\hat g_4$ (\ref{gtrans}) that is explained
in appendix~\ref{appa}.

The cylinder diagram can be expressed as a closed-string propagator
between the appropriate boundary states The overlap between a boundary
state of a euclidean $Dp$-brane located at a transverse position
$y_1^I$ and a euclidean $Dp'$-brane at $y_2^I$ is given by \cite{bgg},
\be
\A_{ p' \eta'; p \eta} (t; {\bf y_1}, {\bf y_2}  )=
 \langle\!\langle p',{\bf y_2},\eta'  |\!| e^{-2 \pi t H^{closed} p^+}
|\!| p, {\bf y_1},\eta \rangle\!\rangle\,,
\label{matclosed}
\ee
where $t=X^+/2\pi p^+$, $H^{closed}$ is the closed-string light-cone
gauge hamiltonian, and $\eta, \eta' =+$ indicates that the boundary
state describes an instanton, whereas $\eta, \eta'=-$ indicates an
anti-instanton. The overlap can be determined by standard methods
(noting that the `in'-state and the `out'-state
have opposite light-cone momentum $p^+$, and therefore $m$
takes the opposite value for the `in'- and `out'-state).

Let us begin by discussing the case of the overlap between two
$(0,0)$-branes. As we have seen the fermionic
ground state (\ref{boundzero}) is an eigenstate of the closed string
light cone Hamiltonian, and therefore the overlap between two
$(0,0)$-branes vanishes so that
$\A_{ (0,0); (0,0)} (t; {\bf y_1}, {\bf y_2}) =0$.
The same comment also applies to the overlaps
between the  $(0,0)$ and $(3,1)$ (or $(1,3)$) boundary states,
$\A_{ (0,0); (3,1)} (t; {\bf y_1},{\bf y_2})=0$. Both of these results
are direct consequences of the corresponding statements in the
flat-space case \cite{gg}.

Next consider the case of the overlap between a $(0,0)$ and a
$\overline{(0,0)}$. Using the boundary state (\ref{boundone}) and the
fact that
$R^\pm(m,\eta) R^\pm(-m,-\eta)= (R^\pm(m,\eta))^2$, together with
standard oscillator algebra leads to the expression for the cylinder,
\be
\A_{\overline{(0,0)}  ;(0,0) } (t;  {\bf y_1},{\bf y_2}) =
h_0({\bf y}_1,{\bf y}_2)\,
{\left(g^{(m)}_2(t)\right)^4\over
\left(f_1^{(m)}(t)\right)^8}\, ,
\label{closeddef}
\ee
where $f_1^{(m)}$ arises from each of the transverse integer-moded
bosons (and is defined in (\ref{fdef}), see \cite{bgg}) and
\be
h_0({\bf y}_1,{\bf y}_2) =
\exp\left( - {m \, (1+q^m) \,
({\bf y}_1^2+{\bf y}_2^2) \over 2\, (1-q^m)}
+ {2 \, m \, q^{m\over 2}
{\bf y}_1\cdot  {\bf y}_2 \over (1-q^m)} \right) \,.
\label{h0}
\ee
Furthermore, each pair of fermions gives a factor
\be
g_2^{(m)}(t) = 4\pi {m}\, q^{-2\Delta_m}\, q^{m/2}\,
\prod_{n=1}^{\infty}
\left( 1 + \left(\omega_n+m \over \omega_n-m\right)
q^{\omega_n} \right)
\left( 1 + \left(\omega_n-m \over \omega_n+m\right)
q^{\omega_n} \right)\,,
\label{g2deff}
\ee
where the `offset' $\Delta_m$ is the same offset that arises in the
definition of $f_1^{(m)}$. In particular, the total offset of
(\ref{closeddef}) is therefore $q^{2m}$, in agreement with the fact
that the lowest closed string state that couples to the boundary
states is the fermionic ground state in (\ref{boundzero}) whose
light-cone energy is $2m$. (This state is characterised by the
condition that is is annihilated (for $\eta=+1$, say) by $\theta^a_L$
and $\theta^a_R$.)

Similar arguments for the overlap between a $(0,0)$ and a
$(0,2)$  at the origin lead to
\be
\A_{(0,0);(0,2)} (t;  {\bf y}, {\bf 0} )
= 2 \sinh(\pi m)\, j_0({\bf y})\,
{\left(g^{(m)}_2(t^2)\right)^2\over
\left(f_1^{(m)}(t)\right)^6 \left(f_2^{(m)}(t)\right)^2} \, ,
\label{DinstD1}
\ee
where the prefactor of $2\sinh(\pi m)$ comes from the normalisation of
the $(0,2)$ boundary state that was determined in \cite{bgg}, and
${\bf y}$ describes the position of the $(0,0)$  (the
$(0,2)$  is assumed to be located at the origin in the transverse
space). Furthermore
\be
j_0({\bf y}) =
\exp\left(- {m \, (1+q^m) \, {\bf y}_t^2 \over 2\,(1-q^m)} \right)
\exp\left(- {m \, (1-q^m) \, {\bf y}_l^2 \over 2\,(1+q^m)} \right) \,,
\label{hhat}
\ee
where ${\bf y}_t$ is the component of ${\bf y}$ in the directions
transverse to the $(0,2)$ , while ${\bf y}_l$ denotes the
components of ${\bf y}$ along the world-volume directions of the
$(0,2)$.

Similarly, the overlap of a $(0,0)$
with a class I $(2,4)$-brane (a euclidean
$D5$-brane) is
\be
\A_{(0,0);(2,4)} (t; {\bf y}, {\bf 0} ) = {j_0({\bf y}) \over
2\sinh(m\pi)} \,
{\left(g^{(m)}_2(t^2)\right)^2\over
\left(f_1^{(m)}(t)\right)^2 \left(f_2^{(m)}(t)\right)^6} \,,
\label{DinstD5}
\ee
where now ${\bf y}_t$ is the component of ${\bf y}$ in the directions
transverse to the $(2,4)$-brane, while ${\bf y}_l$ denotes the
components of ${\bf y}$ along the world-volume directions of the
$(2,4)$.

\section{The open string point of view}
\label{openstring}

In this section the $(0,0)$ ($D$-instanton) is analysed from the open
string point of view. It follows from the equations of motion
\cite{dp} that the open string functions satisfy the equations
\begin{eqnarray}
\partial_+ S(\sigma,\tau) & = & \hm\, \Pi\,
\tilde{S}(\sigma,\tau) \,, \\
\partial_- \tilde{S}(\sigma,\tau) & = &
- \hm\, \Pi\, S(\sigma,\tau) \,,
\end{eqnarray}
where $\hm=\mu x^+$ is the mass parameter in the open string light
cone gauge for which both light cone directions satisfy Dirichlet
boundary conditions \cite{bgg}. The general solution can be written as
\cite{bpz}
\begin{eqnarray}\label{modeexpansion}
S(\sigma,\tau) & = & S_0' \cos(\hm\tau) + \Pi \tilde{S}_0' \sin(\hm\tau)
+ S_0 \cosh(\hm\sigma) + \Pi \tilde{S}_0 \sinh(\hm\sigma)
\nn\\
& & \qquad + \sum_{n\ne 0} c_n \left[
S_n\,  e^{-i(\hom_n\tau - n \sigma)}
+   {i\over \hm} (\hom_n-n) \Pi \tilde{S}_n
        e^{-i(\hom_n\tau + n \sigma)} \right] \,, \\
\tilde{S}(\sigma,\tau) & = &
-\Pi S_0' \sin(\hm\tau) + \tilde{S}_0' \cos(\hm\tau)
+ \tilde{S}_0 \cosh(\hm\sigma) + \Pi S_0 \sinh(\hm\sigma)  \nn\\
& & \qquad + \sum_{n\ne 0} c_n \left[
    \tilde{S}_n e^{-i(\hom_n\tau + n \sigma)}
- {i\over \hm} (\hom_n-n) \Pi S_n e^{-i(\hom_n\tau - n \sigma)}
\right]\,,
\end{eqnarray}
where $\hom_n=\sqrt{\hm^2+n^2}$.

\subsection{Boundary conditions and mode expansions}

The boundary conditions on the open string determine the mode
expansion.  The nature of this expansion depends sensitively on the
particular branes on which the open string terminates and we will
discuss them on a case by case basis.

\subsubsection{$(0,0)$ --- $(0,0)$}

In the case of the open string between a $(0,0)$-brane at
${\bf y}_1$, and a $(0,0)$-brane at ${\bf y}_2$
the open-string bosonic coordinates have the mode expansion
\ba
x^I & = & y_1^I\cosh(\hm\sigma)
+ {y_2^I -y_1^I \cosh(\hm\pi)\over \sinh(\hm\pi)} \sinh(\hm\sigma) +
 \sum_{l\ne 0} {2\over \hom_l} \alpha^I_l
      e^{-i\hom_l\tau} \sin(l\sigma) \,,\nn\\
{\cal P}^I & = & -2i \sum_{l\ne 0} \alpha^I_l e^{-i\hom_l\tau}
          \sin(l\sigma) \,,\label{bosonmode}\\
x^{\prime I} & = &  \hm y_1^I \sinh(\hm\sigma)
+ \hm\, {y_2^I -y_1^I\cosh(\hm\pi)\over \sinh(\hm\pi)} \cosh(\hm\sigma)
 +2  \sum_{l\ne 0} {l\over \hom_l}
       \alpha^I_l e^{-i\hom_l\tau} \cos(l\sigma) \,.\nn
\ea
On the other hand the fermionic modes  are restricted by
the boundary conditions
\be\label{openbound1}
S(\sigma,\tau)=\tilde{S}(\sigma,\tau) \qquad \hbox{for $\sigma=0,\pi$}
\,.
\ee
For the zero-modes this condition requires that $S_0' =
\tilde{S}_0' =0$  and that the second set of zero modes
are related,
\be
\label{zeromodes}
S_0 = \tilde{S}_0\,.
\ee
Furthermore, (\ref{openbound1}) implies that the non-zero fermion modes of
(\ref{modeexpansion}) are related by
\be\label{moderelation1}
\tilde{S}_n = T_n S_n \,,
\ee
where $n\in\Zop$ with $n\ne 0$ and
\be\label{Tdefinition}
T_n = {1\over \hom_n}
       \left[n \bbbone + i \hm \Pi \right] \,.
\ee
The matrix $T_n$ is unitary, $T_n T_n^\ast = \bbbone$.

We can now discuss the open-string supercharges.
The mode expansions of $S$ and $\tilde{S}$, together with
(\ref{moderelation1}) lead to the expansions
\ba
(S -\tilde{S}) & = & 2 \sum_{n\ne 0} c_n
\sin(n\sigma) \left( i \bbbone -
{(\hom_n - n) \over \hm}\, \Pi \right) S_n e^{-i\hom_n\tau}
 \,,\nn \\
(S+\tilde{S}) & = & 2 S_0 \cosh(\hm\sigma) + 2 \Pi S_0 \sinh(\hm\sigma)
\nn\\
& & + 2 \sum_{n\ne 0} c_n \left[
\cos (n\sigma) \left( {n\over \hom_n}
+ i {n\,(\hom_n-n)\over \hm\, \hom_n}\, \Pi \right) S_n \right.
\label{sexpans} \\
& & \qquad\qquad \left.
+ i \sin (n\sigma) \left( {(\hom_n-n)\over\hom_n}
- i {\hm\over \hom_n}\, \Pi \right) S_n  \right]
e^{-i\hom_n\tau} \,. \nn
\ea
The combination $S(\sigma)+\tilde S(\sigma)$ is
proportional to the density of the kinematic supercharge.  Since this
contains a term proportional to
$\sum_{n\ne 0} \sin(n\sigma) c_n M_n S_n e^{-i\hom_n \tau}$, the
supercharge, which is  the integral of this density, has a complicated
time dependence and is not conserved.\footnote{The `conserved'
kinematical supercharges for the class I branes are not strictly
speaking independent of $\tau$. However, their $\tau$ dependence is
simply a consequence of the commutation relations of the fermionic
zero modes with the hamiltonian (\ref{hamilcom}).}

The dynamical supercharge of this open superstring is given by the
{\it difference} of the left and right moving supercharges.  In flat
space this can
be seen by a careful analysis of T-duality from the open string on
the $D9$-brane. Using the formulae given in \cite{mt} one has
\be\label{super}
{\cal Q} = {1\over 2\,\sqrt{X^+}}
\int_0^\pi d\sigma
\left[ {\cal P}^I \gamma^I (\tilde{S} -S) +
x^{\prime I} \gamma^I (S+\tilde{S})
+ \hm x^I \gamma^I \Pi (S+\tilde{S}) \right] \,.
\ee
This is time independent (and therefore
conserved) as  is shown in appendix~\ref{dynsusy}, where it is found
to have the mode expansion
\ba
{\cal Q} & = & {2 \over \sqrt{X^+}}\, \Biggl[ y_2^I \gamma^I
\left(\cosh(\pi\hm) S_0 + \sinh(\pi\hm) \Pi S_0 \right)
- y_1^I \gamma^I S_0  \nn\\
& & \qquad \qquad \left.
+ \pi \sum_{n\ne 0} c_n\, \alpha^I_n\, \gamma^I \,
\left(\bbbone - i {(\hom_n -n) \over \hm} \, \Pi \right) S_{-n}
\right]\,.
\label{superex}
\ea
The anti-commutation relations for the fermionic modes are also
determined in appendix~\ref{susyanti}  and are given by
\ba
\{ S^a_n, S^b_m \} & = & \delta^{ab}\,\delta_{n,-m} \qquad \hbox{if
$n\ne 0$ or $m\ne 0$,} \label{fermioncom1}\\
\{ S^a_0, S^b_0 \} & = & {\pi \hm \over 2\, \sinh(\pi\hm)}
\left( \cosh(\pi\hm) \delta^{a,b} - \sinh(\pi\hm) \Pi^{a,b} \right)
\,. \label{fermioncom2}
\ea
It is easy to calculate the anti-commutator of the
supercharges ${\cal Q}_{\dot a}$, and one finds
\be
\{ {\cal Q}_{\dot a}, {\cal Q}_{\dot b} \} = 2 \,
\delta_{\dot a\dot b}\, H^{open}
+ {2\, \pi\,\hm \over X^+}  \,
(\gamma^I \Pi \gamma^J)_{\dot a\dot b}
\left( y_2^I y_2^J - y_1^I y_1^J \right)\,,
\label{anticommsu}
\ee
where the open string Hamiltonian is defined by
\be
{X^+\over 2\pi}\, H^{open} = {\hm \over 2 \sinh(\pi\hm)}
\left(\cosh(\pi\hm) (y_2^I y_2^I + y_1^I y_1^I) - 2 y_2^I y_1^I\right)
+ 2 \pi \sum_{n> 0} \left( \alpha^I_{-n} \alpha^I_n +
\hom_n S^a_{-n} S^a_n \right) \,,
\label{openham}
\ee
and $n\in \Zop$.
The second term in (\ref{anticommsu}) describes a central charge since
it commutes with the supercharges, as well as the rotation generators
in $SO(4)\times SO(4)$.  It is important that there is no mass
term for the fermionic zero modes in this hamiltonian.

Although the expression (\ref{openham}) has been obtained for the open
string joining two $(0,0)$-branes it has the same structure for
strings joining any pair of class II branes.  In particular, there is
no mass term for the fermionic zero modes.  Recall that in the case of
the hamiltonian of the class I $D$-branes the mass term has the form
$S_0 M \Pi S_0$ but such a term vanishes identically for any of the
class II $D$-branes.

\subsubsection{$(0,0)$ --- $\overline{(0,0)}$}
\label{dbard}

Now consider the open string with one end on the $(0,0)$ and
the other on the $\overline{(0,0)}$ (anti $D$-instanton).
The bosons are still described by
(\ref{bosonmode}), but now the boundary conditions for the fermions,
replacing (\ref{openbound1}), is
\be\label{openbound2}
S(0,\tau)=\tilde{S}(0,\tau) \qquad
S(\pi,\tau)=-\tilde{S}(\pi,\tau)\,.
\ee
This boundary condition is incompatible with the presence of
zero modes, which are therefore absent. The boundary condition at
$\sigma=0$, requires that the non-zero modes have to be related as in
(\ref{moderelation1}), while the condition at $\sigma=\pi$ implies
that
\be\label{moderelation2}
\tilde{S}_n = -e^{2\pi i n}\, T_n^\ast\, S_n \,, \qquad \hbox{for
$n\ne 0$.}
\ee
In order to construct a simultaneous solution to
(\ref{moderelation1}) and (\ref{moderelation2}) let us write these
equations in terms of the plus-minus components introduced before in
(\ref{plusminus}). Then (\ref{moderelation1}) becomes
\be
\tilde{S}_n^\pm = T_n^{\pm} S^\pm_n \,,
\qquad \hbox{where} \qquad
T_n^{\pm} = {(n\pm i\hm) \over \hom_n} \,,
\ee
while (\ref{moderelation2}) is
\be\label{moderelation2p}
\tilde{S}_n^\pm = - e^{2\pi i n} \, \left(T_n^{\pm}\right)^\ast\,
S^\pm_n \,,
\ee
and again both equations only hold for $n\ne 0$. It then follows that
there exists a non-trivial solution provided that the mode number $n$
of $S_n^+$ and $\tilde{S}_n^+$ satisfies the transcendental equation
\be
n\in {\cal P}_+ \,:  {n+i\hm \over n-i\hm} = - e^{2\pi i n} \,,
\label{classonep}
\ee
while the corresponding condition for the mode number of $S_n^-$ and
$\tilde{S}_n^-$ is
\be
n\in {\cal P}_- \,:  {n-i\hm \over n+i\hm} = - e^{2\pi i n} \,.
\label{classtwop}
\ee
Both of these spaces of solutions contain a trivial zero mode solution
with $n=0$, but we have seen that this  is not an allowed mode of the
open string. It is obvious that there are infinitely many other
solutions for $n$ in both cases.  For small $\hm$, the solutions to
both equations are close to all the half-odd integers. Also, if
$n\in{\cal P}_+$, \ie\ if $n$ is a solution to the first equation,
then $-n\in{\cal P}_+$, and similarly for ${\cal P}_-$. The
hamiltonian for this open string is again given by (\ref{openham}),
except that in the second sum $n\in{\cal P}_{\pm}$.

\subsubsection{$(0,0)$ --- $(r,r+2)$}
\label{dpd}

We may also analyse the modes of an open string stretching between a
$(0,0)$ and one of the  $(r,r+2)$-branes (or class I branes) of
\cite{bp,bgg}. Let us first discuss the bosonic modes. For the
directions that satisfy a Dirichlet boundary condition at both
end-points (\ie\ the DD directions), the bosons are still described by
(\ref{bosonmode}), and their contribution to the hamiltonian is given
by the first term in (\ref{openham}). For the DN directions, the
hamiltonian also depends on the position of the $(0,0)$. This is a
consequence of the fact, explained in \cite{bgg}, that $x^-$ depends
on the position on the world-volume of the class I branes.
Taking into account the Wick rotation for $x^+$ that is described in
\cite{bgg}, it follows that the relevant contribution to the open
string hamiltonian is given by
\be
H^{open}_{0} = {\hm\, \tanh(\hm\pi)\over 2} {\bf y}_l^2 \,.
\ee
In order to describe the fermions let us assume the $(r,r+2)$-brane
is at $\sigma=0$, while the $(0,0)$  is at
$\sigma=\pi$. The former boundary condition
($S(0,\tau)=M\tilde{S}(0,\tau)$) then leads
to
\be\label{boundhalf}
S_n = M \tilde{S}_n\,, \qquad n\ne 0
\ee
as well as
\be\label{boundhalfzero}
S_0 = M \tilde{S}_0\,, \qquad
S_0' = M \tilde{S}'_0 \,,
\ee
where $-\bbbone = \Pi M \Pi M$ \cite{dp}. At $\sigma=\pi$
the relevant boundary condition is $S(\pi,\tau)=\tilde{S}(\pi,\tau)$;
this requires that the condition
\be\label{moderelation3}
\tilde{S}_n = e^{2\pi i n}\, T_n\, S_n \,, \qquad n\ne 0
\ee
is satisfied, as well as
\be\label{boundinsthalfzero}
S_0 = \tilde{S}_0\,, \qquad
S_0'=\tilde{S}'_0=0\,.
\ee
As before (\ref{moderelation3}) can be rewritten as
\be\label{moderelation3p}
\tilde{S}_n^\pm =  e^{2\pi i n} \, T_n^{\pm}\,
S^\pm_n \,, \qquad n\ne 0\,.
\ee
The open string modings now depend sensitively on the symmetry of
$M$:

\noindent {\bf Case (i)} $M$ is symmetric and anti-commutes with
$\Pi$. This is the case for the $(1,3)$ and $(3,1)$-brane where, for
example,
$M=\gamma^1\gamma^2\gamma^3\gamma^5$. Since $M^2=\bbbone$
half of the zero modes $S_0$ (namely those that have eigenvalue $+1$
under the action of $M$) satisfy both (\ref{boundhalfzero}) and
(\ref{boundinsthalfzero}). The open string therefore contains four
fermionic zero modes (that commute with the light cone hamiltonian).
As regards the non-zero modes, (\ref{boundhalf}) can be rewritten in
terms of the $\pm$ components as
\be
{S_n^+ \choose S_n^-} = \pmatrix{ 0 & M \cr M^T & 0}
{\tilde{S}_n^+ \choose \tilde{S}_n^-} \,.
\ee
Combining this equation with (\ref{moderelation3p}) gives
\be
{S_n^+ \choose S_n^-} =  e^{2\pi i n}
\pmatrix{ 0 & M T_n^{-} \cr
M^T T_n^{+} & 0}
{S_n^+ \choose S_n^-} \,.
\ee
The condition for a consistent boundary condition is therefore
\be
S_n^+ = e^{4\pi i n} \, T_n^{-} \,
T_n^{+} \,M\, M^T\, S^+_n\,.
\ee
Since $T_n^{+}\, T_n^{-} = 1$, this is the same condition as in flat
space, and thus the modings of these open strings are
exactly as in the flat space case. Since the number of DD and ND
directions of the $(3,1)$-brane are both equal to four,
four of the eight fermions are integer
moded while the other four are half-integer moded.

\noindent {\bf Case (ii)} $M$ is anti-symmetric and commutes with
$\Pi$ which is the case for the $(2,4)$, $(4,2)$, $(2,0)$ and $(0,2)$.  One
example is the D1-brane with $M=\gamma^1 \gamma^2$. In this case,
$M^2=-\bbbone$, and therefore none of the zero modes $S_0$ has
eigenvalue $+1$ under the action of $M$. In particular, it
therefore follows that there are no fermionic zero modes in this open
string.

As regards the non-zero modes, (\ref{boundhalf}) can be
rewritten as
\be
{S_n^+ \choose S_n^-} = \pmatrix{ M^+ & 0 \cr 0 & M^-}
{\tilde{S}_n^+ \choose \tilde{S}_n^-} \,.
\ee
The eigenvalues of $M^\pm$ are $+i,+i,-i,-i$. Let us consider one
of the plus-components $S_n^+$  (the analysis for the minus
components is analogous). The conditions obtained from
(\ref{boundhalf}) and (\ref{moderelation3p}) then become
\be
1 = \pm i e^{2\pi i n} {1 \over \hom_n}(n+i\hm) \,,
\ee
where the sign on the right hand side depends on whether the
eigenvalue under $M^+$ is $\pm i$. Squaring this identity gives
\be
-1 = e^{4\pi i n} {n+i\hm \over n-i\hm} = e^{2\pi i (2n)}
{(2n) + i (2\hm)  \over (2n) - i (2\hm)} \,.
\ee
Thus for each pair of $S^+_n$ modes with eigenvalues $\pm i$, the mode
numbers must satisfy $2n\in {\cal P}_+^{(2\hm)}$. Similarly, for each
pair of $S^-_n$ modes with eigenvalues $\pm i$, the mode numbers must
satisfy $2n\in {\cal P}_-^{(2\hm)}$.

\subsection{Consistency of one-loop open-string amplitudes}
\label{oneloop}

Given the explicit knowledge of the open string spectra involving the
$D$-instanton boundary condition, we can now also evaluate the cylinder
diagram as a trace over the open string states, giving \cite{bgg}
\be
Z_{{p_1;p_2}}(\tilde{t}) =\tr e^{-{X^+\over 2\pi} H^{open}\, \ttt}
\,,\label{matopen}
\ee
where $\ttt = 1/t = 2\pi p^+ / X^+$ and $H^{open}$ is the open-string
hamiltonian in the light-cone gauge. The resulting expressions must
agree with what was obtained in the previous section from a
closed-string point of view.  This is the open-closed consistency
condition that will be checked in the following.

The open string with $(0,0)$-brane  boundary conditions
at both ends, has eight integer-moded world-sheet bosons and eight
integer-moded world-sheet fermions.
Furthermore, the fermionic zero modes commute
with the Hamiltonian, and the trace therefore vanishes, in agreement
with the closed string result.

The analysis is similar for the open string with a $(0,0)$
boundary condition at one end, and a $(1,3)$ boundary condition at
the other. Again, there are four fermionic zero modes that commute
with the open string hamiltonian, and thus the cylinder diagram
vanishes, in agreement with the closed string result.

The situation is more interesting for the case of the open string with
one boundary on the $(0,0)$, and the other on the
$\overline{(0,0)}$. Evaluating the trace now gives an expression of the
form
\be
Z_{(0,0);\overline {(0,0)}} =  \hat{h}_0({\bf y}_1,{\bf y}_2)\,
{\left(\hg^{(\hm)}_4(\tilde t)\right)^4\over
\left(f_1^{(\hm)}(\tilde t)\right)^8}\,,
\label{opendef}
\ee
where $\hat{h}_0({\bf y}_1,{\bf y}_2)$ describes the contribution from
the first term in (\ref{openham})
\be
\hat{h}_0({\bf y}_1,{\bf y}_2) = \exp\left(
- {\hm\,\ttt \over 2 \sinh(\pi \hm)} \left[
\cosh(\pi\hm) ({\bf y_1}^2+{\bf y}_2^2)
- 2 {\bf y}_1 \cdot {\bf y}_2 \right] \right) \,,
\label{overlapresult}
\ee
and the function $\hg^{(\hm)}_4(\tilde t)$ is defined by
\be
\hg^{(\hm)}_4(\tilde t) = \tilde{q}^{-\widehat{\Delta}_{\hm}}
\prod_{l\in{\cal P}_+}
\left(1-\tilde{q}^{\,\widehat\omega_l}\right)^{1\over 2}
\prod_{l\in{\cal P}_-}
\left(1-\tilde{q}^{\,\widehat\omega_l}\right)^{1\over 2} \,.
\label{g4hatdef}
\ee
The products in (\ref{g4hatdef}) are over all the values of $l$ that
satisfy (\ref{calpplusdef}) and (\ref{calpmindef}),
respectively. In both cases, the value $l=0$ is included in the
product. The total $l=0$ contribution $(1-\tq^{\,\hm})^4$ cancels the
zero mode contribution from $(f_1^{(\hm)}(\tilde t))^8$. This is in
agreement with the spectrum since the contribution of the bosonic zero
modes is already described by the prefactor
$\hat{h}_0({\bf y}_1,{\bf y}_2)$. The off-set $\widehat{\Delta}_{\hm}$
is defined in (\ref{deltsdefs}) of appendix~\ref{appa}.

Using the relations $q^{m}=e^{-2\pi t m } = e^{-2\pi \hm}$, as well as
$m=\hm \ttt$, it is easy to see that
$h_0({\bf y}_1,{\bf y}_2)=\hat{h}_0({\bf y}_1,{\bf y}_2)$.
Furthermore, as is shown in appendix~\ref{modprops}, the functions
$g_2$ and $\hg_4$ satisfy the non-trivial identity
\be
g^{(m)}_2(t) = \hg^{(\hm)}_4(\tilde t) \,.
\label{gtrans1}
\ee
This implies that
\be
\A_{p_1;p_2} (t) = Z_{p_1;p_2}(\tilde t)\,,
\label{openclosed}
\ee
and thus that the two calculations agree, as they should.

Similarly, the one-loop contribution of an open string
between the $(0,0)$ and the $(2,0)$ is
\be
Z_{(0,0); (2,0)} =  \hat{\jmath}_0({\bf y})\,
\tilde{q}^{-{\hm\over 2}} \, (1-\tilde{q}^{\,\hm})\,
{\left(\hg^{(2\hm)}_4(\tilde t^{1/2})\right)^2\over
\left(f_1^{(\hm)}(\tilde t)\right)^6
\left(f_4^{(\hm)}(\tilde t)\right)^2}\,,
\label{openDinstD1}
\ee
where the prefactor of $\tilde{q}^{-\hm/2} (1-\tilde{q}^{\,\hm})$ cancels
the part of the zero mode contribution from $(f_1^{(\hm)}(\ttt))^6$
that is not cancelled by $(\hg^{(2\hm)}_4(\tilde t^{1/2}))^2$, and
\be
\hat{\jmath}_0({\bf y}) = \exp\left(
- {\hm\,\ttt \over 2}
\left[ {\bf y}_t^2 {\cosh(\pi\hm)\over \sinh(\pi\hm)}
+ {\bf y}_l^2 {\sinh(\pi\hm)\over \cosh(\pi\hm)} \right] \right)\,.
\ee
Again, it is easy to see that $j_0({\bf y})=\hat{\jmath}_0({\bf y})$,
and that (\ref{openclosed}) is again a consequence of
(\ref{gtrans1}). Finally, the result for the case of the open string
between the $(0,0)$ and the $(4,2)$-brane is
\be
Z_{(0,0); (4,2)} =  \hat{\jmath}_0({\bf y})\,
\tilde{q}^{\,{\hm\over 2}} \, (1-\tilde{q}^{\,\hm})^{-1}\,
{\left(\hg^{(2\hm)}_4(\tilde t^{1/2})\right)^2\over
\left(f_1^{(\hm)}(\tilde t)\right)^2
\left(f_4^{(\hm)}(\tilde t)\right)^6}\,,
\label{openDinstD5}
\ee
and the consistency of the closed-string and open-string sectors
follows by the same arguments as before.

\section{$(4,0)$ and $(0,4)$ with flux}
\label{zerfour}

The $(4,0)$-brane and the $(0,4)$-brane couple to the self-dual
background \RR\ four-form potential.  This means that a nontrivial
Born--Infeld flux is necessarily switched on in the world-volume.  In
turn, this affects the Neumann boundary conditions.  We will determine
the open-string boundary conditions by enforcing the consistency of
the cylinder diagrams under the $S$ modular transformation, starting
from the closed-string boundary states.

For definiteness, let us consider the $(4,0)$-brane  (the construction
for the $(0,4)$-brane is similar).   The corresponding boundary state
should be characterised by the gluing conditions
\be
\left( Q_{\dot a} + i\,\eta\, \Pi_{\dot a\dot b}\,
{\tilde Q}_{\dot b}\ \right)
|\!| (4,0),\eta\,\rangle\!\rangle = 0\,.
\label{dynamical4}
\ee
In terms of the chiral $\pm$ components this is the condition
\be
\left( Q^{\pm}_{\dot a} \pm i\,\eta\, {\tilde Q}^\pm_{\dot a}\ \right)
|\!| (4,0),\eta\,\rangle\!\rangle = 0\,.
\label{dynamical4pm}
\ee
In addition to (\ref{dynamical4pm}), the boundary
state should satisfy the Dirichlet gluing conditions
\ba
\left( \alpha^{i'}_k - \tilde\alpha^{i'}_{-k} \right)
|\!| (4,0),\eta\,\rangle\!\rangle  &=& 0\,,
\qquad k\in\Zop\setminus\{0\}\,, \nn\\
\left( \bar{a}^{i'}_0 - a^{i'}_0 + i \sqrt{2m} y_t^{i'}\right)
|\!| (4,0),\eta\,\rangle\!\rangle & = & 0 \,,
\label{boundarycond4prime}
\ea
where $i'$ denotes the coordinates that are transverse to
the $(4,0)$   and $y_t^{i'}$ is the position of the
$(4,0)$  in these directions.  As before,
(\ref{dynamical4pm}) together with (\ref{boundarycond4prime})
already determines the gluing conditions for all fermionic
modes. Once these have been determined they will imply, using
(\ref{dynamical4pm}), what the gluing conditions for the bosonic modes
along the world-volume of the $(4,0)$  must be.

{}From the condition that $(Q+i\eta \tilde{Q})^+$ should
annihilate the boundary state, using
(\ref{boundarycond4prime}), we find that
\be\label{fermion40p}
\left(S^+_n + i \eta {\omega_n - m\,\eta \over n}
\tilde{S}^+_{-n} \right)
|\!| (4,0),\eta\,\rangle\!\rangle = 0  \,, \qquad n\ne 0 \,,
\ee
while the condition that $(Q-i\eta \tilde{Q})^-$ annihilates the boundary
state gives
\be\label{fermion40m}
\left(S^-_n - i \eta {\omega_n - m\,\eta \over n}
\tilde{S}^-_{-n} \right)
|\!| (4,0),\eta\,\rangle\!\rangle = 0  \,, \qquad n\ne 0 \,.
\ee
Both of these identities follow directly from the analysis described
before for the ${(0,0)}$-brane. With these fermionic gluing
conditions, the terms in (\ref{dynamical4pm}) proportional to
$\alpha^{i'}_n$ annihilate the boundary state. On the other hand, the
terms that are proportional to $\alpha^{i}_n$ vanish if and only if
\be\label{boson40m}
\left[ \alpha^i_{n} +
\left({\omega_n - m\,\eta \over \omega_n + m\,\eta} \right)
\tilde\alpha^i_{-n} \right]
|\!| (4,0),\eta\,\rangle\!\rangle = 0
\ee
for all $n\ne 0$. This condition reduces to the standard Neumann
boundary condition for $m\rightarrow 0$.

The zero-mode
component of (\ref{dynamical4pm}) requires, using the second equation
of (\ref{boundarycond4prime}), that
\be\label{fermion400}
\left( S_0^a + i \eta\,\Pi_{ab} \tilde{S}_0^b \right)
|\!| (4,0),\eta\,\rangle\!\rangle = 0 \,.
\ee
Furthermore, the bosonic zero-mode condition for the transverse
directions is
\be\label{boson400}
\left( p_0^i - i\, \eta\,m \, x_0^i\right)
|\!| (4,0),\eta\,\rangle\!\rangle =
0\,.
\ee

So the complete set of gluing conditions
for the supersymmetric $(4,0)$-brane with flux is
given by  (\ref{boundarycond4prime})-(\ref{boson400}).
The bosonic gluing conditions can be summarised as
\be
\label{bosglues}
\left. \left({\cal P}^i - i \, \eta\, m \,  x^i \right)\right|_{\tau=0}
|\!| (4,0),\eta\,\rangle\!\rangle = 0 \,.
\ee
A notable feature of condition (\ref{boson400}) is that for $\eta = +$
the bosonic zero mode ground state is the Fock space ground
state since it is annihilated by $\bar a^i$. On the other hand,
the ground state for the anti-brane ($\eta=-$) would be the state that
is killed by the zero mode {\it creation} operator, $a^i$.  This would
mean that the anti-brane had to lie in a different Fock space that
decouples from all of the other branes. The conclusion is therefore
that there is no supersymmetric anti-brane for the $(4,0)$ (and
$(0,4)$) cases. Which of the two, the brane or the anti-brane, is
supersymmetric obviously depends on the sign of $m$, and thus on the
sign of the \RR\ background flux.

\subsection{The open string description}

In order to deduce the open-string description of this boundary
condition, let us analyse one of the non-vanishing overlaps involving
the $(4,0)$-brane. To be specific, consider its overlap
with the $(2,0)$-brane, where, for simplicity, both branes are taken
to be at the origin in the transverse space. Using the same arguments
as before, one finds that this overlap is
\be
\A_{(2,0);(4,0)}(t) = (2 \sinh(\pi m))^2 \, {
\left(g_{1,-}^{(m)}(2t)\right)^2\, \left(g_{2,-}^{(m)}(2t)\right)^2
\over
\left(f_1^{(m)}(t)\right)^4 \,
\left(g_{1,-}^{(m)}(t)\right)^2\, \left(g_{2,-}^{(m)}(t)\right)^2 }
\,,
\label{overtwof}
\ee
where the functions $g_{1,-}^{(m)}(t)$ and $g_{2,-}^{(m)}(t)$ and
their transformation properties are defined in appendix~\ref{factorg}.
Using the formulae given there, the relevant open string has a
one-loop partition function given by
\be
Z_{(2,0);(4,0)}(\ttt) =  \tilde{q}^{\,-\hm}\,
(1-\tilde{q}^{\,\hm})^2\, {
\left(\hat{g}_{1,+}^{(2\hm)}(\ttt/2)\right)^2\,
\left(\hat{g}_{4,+}^{(2\hm)}(\ttt/2)\right)^2
\over
\left(f_1^{(\hm)}(\ttt)\right)^4 \,
\left(\hat{g}_{1,+}^{(\hm)}(\ttt)\right)^2\,
\left(\hat{g}_{4,+}^{(\hm)}(\ttt)\right)^2 }
\,.
\label{onepart}
\ee
This one-loop amplitude is consistent with the boundary condition
for the $(4,0)$-brane
\be
\tilde{S}^\pm_n = \pm {n+i\hm \over \hom_n} S_n^\pm\,,
\label{consisf}
\ee
as well as
\be
\alpha^i_n  =  {n-i\hm \over n+i\hm}\;  \tilde\alpha^i_n \,, \qquad
\alpha^{i'}_n  =  - \tilde\alpha^{i'}_n \,.
\label{consisalp}
\ee
The bosonic relations are equivalent to the condition that
\ba
{x'}^{i} (\sigma,\tau) - \hm x^{i}(\sigma,\tau) & = & 0 \,, \nn\\
x^{i'}(\sigma,\tau) & = & y^{i'}
\label{newboss}
\ea
at the boundary corresponding to the $(4,0)$ brane, while
the fermionic conditions are equivalent to
\be
S(\sigma,\tau) = \Pi \tilde{S}(\sigma,\tau) \,.
\ee
In section~\ref{twotwosusy} we will point out that the modified bosonic 
conditions (\ref{newboss}) also arise in the context of the
$(+,-;4,0)$-brane as described in \cite{st,hk}. As in the case of the
$(0,0)$--$(0,0)$ system the cylinder connecting two $(4,0)$-branes
vanishes, $\calA_{(4,0);(4,0)}(t) = 0$.

\subsection{Supersymmetry}

Having determined the boundary conditions for the $(4,0)$-brane, we
can now deduce the mode expansion of the bosonic and fermionic
fields for the open string both of whose ends lies on a
$(4,0)$-brane. The components of $S^+$ and $\tilde{S}^+$ have the
mode expansions given in  (\ref{sexpans})  while the  components of
$S^-$ and $\tilde{S}^-$ have the same mode expansion as the
components of $S^-$ and $\tilde{S}^-$ for the
$\overline{(0,0)}$. Thus,
\ba
(S +\tilde{S})^- & = & 2 i \sum_{n\ne 0} c_n
\sin(n\sigma) \left( 1 + i {(\hom_n - n) \over \hm}\right)
S_n e^{-i\hom_n\tau}  \,,\nn \\
(S-\tilde{S})^- & = & 2 S_0^- \cosh(\hm\sigma) + 2 S_0^-
\sinh(\hm\sigma) \nn\\
& & + 2 \sum_{n\ne 0} c_n \left[
\cos (n\sigma) {n\over \hom_n}
\left( 1 + i {(\hom_n - n) \over \hm}\right) \right. \nn\\
& & \qquad\qquad\qquad \left.
+ \sin(n\sigma) {\hm\over\hom_n}
\left( 1 + i {(\hom_n - n) \over \hm}\right)
\right] S_n^- e^{-i\hom_n\tau} \,.
\label{splusmin}
\ea
The bosonic fields $x^{i'}$ have an expansion of the form
(\ref{bosonmode}). On the other hand, the mode expansion for the
bosonic fields in the first four directions is\footnote{We thank
Y. Michishita for correcting an error in an earlier version of this
paper. This mode expansion was described before, in a different
context, in \cite{michishita}.}
\ba
x^i & = & (x_0^i + p_0^i\, \tau)\, e^{\hm\,\sigma} \nn\\
& & \qquad +
2 i \sum_{l\ne 0} {l\over \hom_l (l-i\hm)} \alpha^i_l
      e^{-i\hom_l\tau} \cos(l\sigma)
+ 2 i \sum_{l\ne 0} {\hm\over \hom_l (l-i\hm)} \alpha^i_l
      e^{-i\hom_l\tau} \sin(l\sigma) \,,\nn\\
{\cal P}^i & = &  p_0^i \,e^{\hm\,\sigma} +
2 \sum_{l\ne 0} {l\over (l-i\hm)} \alpha^i_l
      e^{-i\hom_l\tau} \cos(l\sigma)
+ 2 \sum_{l\ne 0} {\hm\over (l-i\hm)} \alpha^i_l
      e^{-i\hom_l\tau} \sin(l\sigma) \,,\\
x^{\prime i} & = &  \hm \, (x_0^i + p_0^i \,\tau) e^{\hm\,\sigma} \nn\\
& & \quad - 2 i \sum_{l\ne 0} {l^2\over \hom_l (l-i\hm)}
\alpha^i_l e^{-i\hom_l\tau} \sin(l\sigma)
+ 2 i \sum_{l\ne 0} {\hm l \over \hom_l (l-i\hm)} \alpha^i_l
      e^{-i\hom_l\tau} \cos(l\sigma) \nn\,.
\label{firstx}
\ea
Using these expansions it is straightforward to show that
the positive $SO(4)$ chirality
 component of the difference of the two supercharges (see
(\ref{super}))
\be\label{superp}
{\cal Q}^+ \equiv {1\over 2} (1+\Pi) {\cal Q} =
{1\over 2\,\sqrt{X^+}}
\int_0^\pi d\sigma
\left[ {\cal P}^I \gamma^I (\tilde{S} -S) +
x^{\prime I} \gamma^I (S+\tilde{S})
+ \hm x^I \gamma^I \Pi (S+\tilde{S}) \right]^+ \,,
\ee
and the negative $SO(4)$ chirality
component of the sum of the two supercharges
\be\label{superm}
\bar{\cal Q}^-  \equiv {1\over 2} (1+\Pi) \bar{\cal Q}=
{1\over 2\,\sqrt{X^+}}
\int_0^\pi d\sigma
\left[ {\cal P}^I \gamma^I (S+\tilde{S}) -
x^{\prime I} \gamma^I (S-\tilde{S})
+ \hm x^I \gamma^I \Pi (S-\tilde{S}) \right]^-
\ee
are conserved. Therefore, this system preserves eight of the
dynamical supersymmetries as was the case with the $(0,0)$--$(0,0)$
system. Furthermore, none of the kinematical supersymmetries is
conserved. This follows from the fact that the integrals of the
equations in (\ref{splusmin}) obtain contributions from the non-zero
modes.

\section{Relation to $(2,2)$ world-sheet supersymmetry}
\label{twotwosusy}

In earlier sections we have have constructed  class II $D$-branes
that preserve half of the light-cone gauge dynamical supersymmetries
for the cases $(0,0)$, $(4,0)$ and $(0,4)$. From the point of view of
our construction it seems that none of the other class II  $D$-branes
(\ie\ the cases $(1,1)$, $(2,2)$, $(3,3)$, $(4,4)$) possess
eight unbroken dynamical supersymmetries.  In a separate
approach the $D$-branes that preserve some supersymmetry in a generic
$pp$-wave background with $(2,2)$ world-sheet supersymmetry
\cite{mm} were recently analysed in \cite{hk}.
Apart from some oblique branes (see below), the only supersymmetric
branes that were found in \cite{hk} were the cases $(0,0)$, $(4,0)$ and
$(0,4)$. We would like to explain how our results fit in with
those of \cite{hk}.

The generalised $pp$-wave backgrounds of \cite{mm} can be
expressed in terms of string theories with $(2,2)$ world-sheet
supersymmetry. These backgrounds preserve at least four of the
sixteen dynamical light-cone gauge supersymmetries. More precisely,
the four parameters of the $(2,2)$ world-sheet supercharges are
interpreted as four components of the space-time Killing spinor
identified in \cite{mm} and are parameterized by two complex
constants, $\alpha$ and $\zeta$.  In the plane-wave background of
interest to us, this is only a sub-symmetry of the complete fermionic
symmetry of the background.  In fact, the four components of the
dynamical Killing spinor
transform in a certain spinor representation of the diagonal $SO(4)$
subgroup of $SO(4) \times SO(4)$.

This can be seen from the expression for the spinor as given in
equation (A.5) of \cite{mm}. The piece of the Killing spinor
proportional to $\alpha$ is identified with the spinor state
$\psi=(+1/2,+1/2,+1/2,+1/2)$ which is the bottom state of the
${\bf 8_s}$ spinor representation of $SO(8)$.  The piece proportional
to $\zeta$ is $\chi=(-1/2,-1/2,-1/2,-1/2)$, which is the top state of
${\bf 8_s}$.

Decomposing ${\bf 8_s}$ with respect to the standard
embedding of $SO(4)\times SO(4)$ in $SO(8)$  gives
\be
{\bf 8_s} = \left({\bf 2_+} \otimes {\bf 2_+}\right) \oplus
            \left({\bf 2_-} \otimes {\bf 2_-}\right) \,,
\label{deceight}
\ee
where ${\bf 2_\pm}$ denote the complex two-dimensional spinors of
$SO(4)$ with chiralities $\pm$.  The states $\psi$ and $\chi$ lie in
the first product  together with the states $(+1/2,+1/2,-1/2,-1/2)$
and $(-1/2,-1/2,+1/2,+1/2)$. Decomposing ${\bf 2_+} \otimes {\bf 2_+}$
with respect to the diagonal $SO(4)$, gives
\be
{\bf 2_+} \otimes {\bf 2_+} = {\bf 2_+} \oplus {\bf 1} \oplus {\bf 1}
\,,
\label{diagso}
\ee
where $\psi$ and $\chi$ generate precisely the ${\bf 2_+}$ of the
diagonal $SO(4)$, while the other two states are singlets.
Similarly,  ${\bf 2_-} \otimes {\bf 2_-}$ can be decomposed with respect
to the diagonal $SO(4)$, and it is easy to see that
\be
{\bf 2_-} \otimes {\bf 2_-} = {\bf 2_-} \oplus {\bf 1} \oplus {\bf 1}
\,.
\label{newdiag}
\ee
Thus the four (real) supercharges considered in \cite{mm} transform in
the ${\bf 2}_+$ representation of the diagonal $SO(4)$, and are in
fact uniquely characterised by this property.

Next we want to analyse which of these supersymmetries are preserved
in the presence of a $D$-brane\footnote{We are here discussing the
lorentzian $(+,-;r,s)$-branes in order to compare with \cite{hk}.}.
Let us concentrate on the sixteen
dynamical supersymmetries of the light-cone gauge type IIB theory.
These transform in the
${\bf 8_s}$ of $SO(8)$ and therefore have chirality $+$ with respect
to the $SO(1,1)$ of the light-cone gauge. The dynamical
supersymmetries that are preserved by a brane are precisely those
that are invariant under the action of
$\hat\Gamma=\prod_{i\in\cal N} \gamma^i$, where the product is over the
gamma-matrices associated with the world-volume directions of the
brane. In light-cone gauge, this translates into the statement that
the dynamical supersymmetries in ${\bf 8_s}$ that survive are
those that are invariant under $\Gamma=\prod_{i\in\cal N'} \gamma^i$,
where ${\cal N}'$ is the set of transverse world-volume directions.
This condition selects out precisely half of the eight complex states.

In relating the results of \cite{hk} to our case (where the background
preserves the maximal amount of supersymmetry) one has to bear in mind
two further restrictions made in \cite{hk}:
\begin{list}{(\roman{enumi})}{\usecounter{enumi}}
\item Since a generic background with  $(2,2)$ world-sheet supersymmetry
only possesses the
space-time Killing spinors described above, all the branes found in
\cite{hk} preserve a linear combination of the spinors $\psi$ and
$\chi$. On the other hand, the maximally supersymmetric background
that is considered here, may (and does) possess $D$-branes that
preserve half the dynamical supersymmetries, none of which
lie in the subspace spanned by $\psi$ and
$\chi$.  Since these do not preserve the $(2,2)$ supersymmetry of
\cite{hk} these branes are absent from the analysis of \cite{hk}.
\item In the analysis of \cite{hk} an ansatz is made for $\epsilon_-$
that is only the most general ansatz if the $D$-brane does not
preserve any kinematical supersymmetries (which correspond to
solutions of the homogeneous $\epsilon_-$ equation of \cite{mm}).
\end{list}
Since all the class I branes preserve half the kinematical
supersymmetries, point (ii) above implies that they should not appear in
the analysis of \cite{hk}, and this is indeed the case. Actually,
all class I branes, except for certain $(1,3)$ and $(3,1)$ branes, only
preserve dynamical supersymmetries that lie outside the subspace
spanned by $\psi$ and $\chi$, and therefore do not appear to be
supersymmetric from the analysis of \cite{hk} because of (i).

As regards the class II branes, suppose that $\Gamma$ is a product of
$\gamma_i$ matrices (as is the case for the $(+,-;r,s)$ branes). Then
for each $(r,s)$ in class II, there exists a configuration for which
$\Gamma$ leaves the space spanned by $\psi$ and $\chi$ invariant. In
order to see this one can use the representation (up to a suitable
normalisation)
\be
\gamma^i = (b^i + b^{+i})\,, \qquad i=1,2,3,4\,, \qquad
\gamma^{4+i} = i (b^i - b^{+i})\,, \qquad i=1,2,3,4\,,
\label{suitnorm}
\ee
where $b^i$ and $b^{+i}$, $i=1,2,3,4$, act as step operators on the
four entries of the spinor states
$(\pm 1/2, \pm 1/2, \pm 1/2, \pm 1/2)$. Furthermore, the
supersymmetric class II branes we have found -- namely the $(0,0)$,
$(4,0)$ and $(0,4)$ branes -- do not preserve any kinematical
supersymmetries, and thus the analysis of \cite{hk} is applicable for
them. Our results for these branes are therefore in agreement with the
findings of \cite{hk}.

Finally, the oblique branes that were found in \cite{hk} preserve a two
complex dimensional subspace of the space spanned by $\psi$ and
$\chi$, but probably not half of the supersymmetries that are
present in our cases. Within the context of our analysis it is
therefore not surprising that we have not encountered them. It should
be straightforward to generalise our construction in order to describe
them as well.

We can also use the connection with $(2,2)$ world-sheet
supersymmetry to confirm  the form of the modified Neumann
boundary conditions (\ref{newboss}). The coupling of the
$(+,-;4,0)$-brane to the background five-form field induces
non-zero Born-Infeld flux $F_{+i}$ that is determined in terms of
the superpotential $W$  (see equation (4.14) of \cite{hk}). For the
maximally supersymmetric plane-wave background the superpotential $W$
is quadratic in the transverse coordinates, and $F_{+i}$ is therefore
proportional to $\mu\, x^i$. The open string boundary condition for a
brane in the presence of this Born-Infeld flux is then
\be
(\partial_\sigma x^i + F_{+i}\, \partial_\tau x^+) = 0 \,.
\label{newborn}
\ee
In the usual open string light-cone gauge that is appropriate for
time-like branes $x^+ = p^+ \tau$, and thus (\ref{newborn}) becomes
\be
x^{\prime \, i} - m \, x^i = 0\,,
\label{bouncor}
\ee
where $m = \mu\, p^+$. Taking into account that the analysis of
section~\ref{zerfour} is formulated for euclidean branes, and that the
mass-parameter in the corresponding open string light-cone gauge is
$\hm$ rather than $m$,  (\ref{bouncor})
agrees precisely with the string equation
we found in (\ref{newboss}).

The condition (\ref{newborn}) agrees also with equation (8.12) of
\cite{st}. There it was argued that the derivative of the condition
follows from the dynamics of the $(+,-;4,0)$-brane described by
the sum of the Dirac--Born--Infeld lagrangian and the
Wess--Zumino term. The latter is proportional to $\int dA \wedge F_5$,
where $A$ is the Born--Infeld vector potential, $F_5$ the
background five-form field strength, and the integral is over a
seven dimensional surface that bounds the brane.

\section{The classical $D$-instanton}
\label{classinst}

In this section we will describe the classical supergravity
$D$-instanton solution that corresponds to the boundary state we
constructed in section \ref{boundstate}. We begin by reviewing the
description of the IIB $D$-instanton in flat space.

\subsection{Review of the flat space description}

The euclidean field equations of classical Type  IIB supergravity
theory possess a $D$-instanton solution in which the dilaton and the
Ramond--Ramond (\RR) scalar fields have nontrivial profiles in the
Einstein frame, while all other fields are trivial \cite{ggp}.
The BPS condition relates the \RR\ scalar ($C^{(0)}$) to the dilaton
$\phi$.  The quantity $e^\phi$ satisfies the ten-dimensional  equation
for a scalar Green function,
\be
\nabla_x^2 e^\phi =  2\pi |K| \delta^{(10)}(x-x_0)\, ,
\label{scalareq}
\ee
where $x_0$ is the position of the instanton.  The \RR\ scalar is given
by $d \hat C^{(0)} = de^{-\phi}$, where $d\hat C^{(0)} = i dC^{(0)}$
(the factor of $i$ arising due to the effects of the Wick rotation to
euclidean signature).
The flat space-time solution of this equation is
\cite{ggp}\footnote{This corrects a small numerical mistake in \cite{ggp}.}
\be
e^{\phi^{(10)}} = g +  h(|x-x_0|)\, ,
\label{teninst}
\ee
where $g= e^{-\phi_0}$ is the string coupling constant and
\be
h =  {3 |K|\over \pi^4 |x-x_0|^8}\, .
\label{hdef}
\ee
The function $h$ is simply the euclidean scalar field propagator.
The corresponding \RR\ scalar field is then given by
\be
C^{(0)} = \chi + i {1\over  g +  h(|x-x_0|)}\, ,
\label{rrscal}
\ee
where $\chi$ is the constant value of the field.

The instanton carries a charge $K$ which represents the violation of the
Noether symmetry  associated with the translation symmetry of the \RR\
scalar, $C^{(0)} \to C^{(0)} + b$.   The Noether current is given by
\be
j_\mu =  i\, e^{2\phi}\, \partial_\mu C^{(0)} \,,
\label{currentn}
\ee
and the charge carried by the $D$-instanton is given by the integral of
the radial component of the current over a nine-sphere enclosing the
point $x=x_0$,
\be
q =  \oint d \Sigma_\mu i\, e^{2\phi} \, \partial^\mu C^{(0)}\, ,
\label{chargeviol}
\ee
where $d\Sigma_\mu$ is the element of area on the nine-sphere.
A generalization of Dirac's argument for the quantization of magnetic
charge in the presence of an electric charge leads to a quantization
of $K$ in the presence of a $(4,4)$-brane.

It is straightforward to see that the above solution
carries a charge that is an integer multiple of $2\pi$,
\be
q =  \oint d\Sigma_\mu {24 K (x^\mu - x_0^\mu)\over
\pi^4 |x-x_0|^{10}}  =  2\pi K\, .
\label{totcharge}
\ee
The $D$-instanton action is equal to $2\pi |K|$.

The euclidean solution has an interpretation in lorentzian signature
space-time as a tunneling process in which the \RR\ charge changes by
$K$ units between the initial time $x^0 \to -\infty$ and the final
time $x^0 \to + \infty$. In order to see this, the solution  must be
continued to lorentzian signature with a suitable $i \epsilon$
prescription which reproduces the causal properties of the Feynman
propagator for the non-constant term in (\ref{teninst}).  For our
purposes it is of interest to express the fields in light-cone
coordinates.  In that case the time coordinate is $x^+$ and the
Noether charge of the  \RR\ scalar is $i \partial_+ C^{(0)}$.  We may
write the solution in the form (letting $X = x- x_0$ for convenience)
\be
e^{\phi^{(10)}} = g +{|K| \over 2\pi^4} \int_0^\infty ds\, s^3\,
e^{is\left(-2 X^+ X^- + {\bf X}^2 + i\epsilon\right)} \, .
\label{sollight}
\ee
It is natural to Fourier transform  this solution  with respect to
$X^-$ to express it in terms of the mixed
$(p^+,  X^+, {\bf X})$ representation.  This gives a factor of
$\delta(2s X^+ - p^+)$, which only has support when $p^+/X^+- \ge
0$ and the result is
\be
\widehat{e^{\phi^{(10)}}} \equiv {1\over {2\pi}} \int dX^-\,
e^{\phi^{(10)}} \, e^{ip^+ X^-} = g \delta(p^+)
+ {|K|\, (p^+)^3 \over 2(2\pi X^+)^4} e^{ip^+{\bf X}^2/2X^+}
\, .
\label{fouriersol}
\ee
This becomes a real solution after a conventional
Wick rotation of the light-cone
time variable, $X^+ \to - i X^+$.
In contrast to our earlier
discussion we shall not perform this Wick rotation in this section.
This has to be taken into account when comparing with the results of
section~2.

In light-cone coordinates the charge (\ref{chargeviol}) is the
difference between the final and initial charges, defined at
$X^+ = +\infty$ and $X^+ = -\infty$, respectively
\be
q = q_f -q_i\, .
\label{qdiffs}
\ee
Here $q_f$ and $q_i$ are the charges defined at
$X^+ = X^+ _f = +\infty$ and $X^+ = X^+ _i =-\infty$, respectively.
The expression for $q_f$ is
\be
q_f =\left.
 \int d\Sigma_+\,   {K\over 2 (2 \pi X^+_f)^4} \int_0^\infty ds s^4
e^{is\left(
- X^- + {\bf X}^2/2X^+_f + i\epsilon/2X^+_f\right)}
\right|_{X^+_f=\infty}\, ,
\label{chargeinx}
\ee
where $d\Sigma_+ = d^8 X^i dX^-$.  When expressed in terms of the
Fourier  transformed solution the result is
\be
q_f = \lim_{p^+ \to 0_+} 2\pi \int d^8 X^i\, {K\over 2}  \,
\left({p^+ \over 2\pi X^+_f}\right)^4 \,
e^{ip^+{\bf X}^2/2X^+_f} \,.
\label{fourch}
\ee
Performing the ${\bf X}$ integrations in $d\Sigma_+$ gives
\be
q_f = \pi K \,,
\label{newres}
\ee
where a
Wick rotation of $X^+$
has again been assumed,
which makes the integral convergent. Similarly we have
$q_i = - \pi K$.

\subsection{The classical solution for the plane-wave background}

The solution of (\ref{scalareq}) in a non-trivial conformally flat
geometry with vanishing scalar curvature and metric
$ds^2 = \Omega^2 \,dx^2$ can be straightforwardly written in terms of
the harmonic function $h$ in the form
\be
e^{\tilde \phi} =g +  \Omega^{-4}(x) \Omega^{-4}(x_0)
h(|x-x_0|)\, .
\label{confsol}
\ee
For example, in the case of $AdS_5\times S^5$, $\Omega = \rho^{-1} L$,
where $\rho$ is the radial coordinate
($\rho^2 \equiv \sum_{i=5}^9 (x^i)^2$) and $L$ is the scale of the
$AdS_5$ \cite{bgkr}.
The plane-wave background of interest to us has the metric
\be
ds^2 = -2 dx^+ dx^-  - (\pi\mu)^2 (dx^+)^2 {\bf x}^2 + d \x^2\, ,
\label{ppmetric}
\ee
where $\x$ denotes the eight transverse coordinates. This  may be
expressed in conformally flat coordinates by transforming to Rosen
coordinates (denoted by tilde's)
\be
\tilde x^+ =  {1\over \pi\mu} \tan \pi\mu x^+\, , \qquad
\tilde \x = {\x \over \cos\pi \mu x^+}\, ,\qquad
\tilde x^- =  x^- + {\pi\mu \over 2} \tan \pi\mu x^+ \x^2\, .
\label{rosent}
\ee
In this system of coordinates the metric becomes
\be
ds_{Rosen}^2 = (1 + (\tilde x^+)^2)^{-1} (-2
d \tilde x^+ d\tilde x^- + d\tilde
{\bf x}^2)\, ,
\label{metrosen}
\ee
and thus the conformal prefactor is
$(1 + (\tilde x^+)^2)^{-1} = \cos^2 \pi\mu x^+$.
The invariant finite squared length element is therefore
\be
\Phi_{Rosen} = \cos \pi\mu x^+ \cos \pi\mu x_0^+ (- 2
 \tilde X^+ \tilde X^- +
{\bf \tilde X}^2)\, .
\label{rosprop}
\ee
In these coordinates the initial and final times  are
$\tilde x^+ = \pm\infty$, which are at the points where
$\cos \pi\mu x^+ = 0$.

It is easy to transform to the global coordinates (\ref{ppmetric})
by substituting the $\tilde x$ variables in terms of $x$,
giving
\be
\Phi = -{2\over \pi\mu} X^- \sin (\pi\mu X^+) - 2 (\x^2+ \x^2_0)
\sin^2\left( {\pi\mu X^+\over 2}\right) + {\bf X}^2\, .
\label{origphi}
\ee
 The dilaton profile
in the background of a $D$-instanton is therefore,
\be
e^{ \tilde \phi} =g + {3|K| \over \pi^4}
{1\over (\Phi(x,x_0) + i\epsilon)^4} \, .
\label{dinpp}
\ee
Apart from the constant, $g$, this is  the expression given for the
scalar Feynman propagator in \cite{mss}.
Fourier transforming with respect to $X^-$ now gives
\be
\widehat{e^{\tilde\phi}} = g \delta(p^+) +
{|K|\, \mu^4 \, (p^+)^3 \over 2 (2\sin(\pi \mu X^+))^4}
e^{i  {\pi\mu p^+ \over 2} \left((\x^2 + \x_0^2)\cot(\pi\mu X^+)
- 2\x\cdot \x_0 /\sin(\pi\mu X^+)\right)}\,.
\label{instglobal}
\ee
The expression for the charge at a positive value
of $X^+$ due to a $D$-instanton at the origin is then
\be
q_f = \lim_{p^+ \to 0_+} 2\pi \int d^8 {\bf x^i} {K\over 2}
\left({\mu p^+  \over 2 \sin (\pi\mu  X^+)}\right)^4
e^{ip^+ \pi\mu {\bf x}^2 \cot (\pi\mu X^+_f)/2}
= \pi K  (\cos (\pi\mu X^+_f))^{-4}\,,
\label{mixppinstf}
\ee
while at negative $X^+$ it is
\be
q_i = \lim_{p^+ \to 0_-} 2\pi \int d^8 {\bf x^i} {K\over 2}
\left({p^+ \mu \over 2\sin (\pi\mu  X^+)}\right)^4
e^{ip^+ \pi\mu {\bf x}^2 \cot (\pi\mu X^+_i)/2}
= -\pi K  (\cos (\pi\mu X^+_i))^{-4} \,.
\label{mixppinsti}
\ee
The dependence of the charge $q_f-q_i$ on $X^+_i$ and $X^+_f$ is a
consequence of  the fact that the \RR\ scalar is not massless.
It would be interesting to understand the properties of the $D$-instanton
in more detail, particularly its  relation to
the Yang--Mills instanton via the duality conjectured in \cite{maldetal}.

\subsection{Supergravity approximation of the interaction energy}

The above discussion implies that one can think of the $D$-instanton
at $x_0$ as a source of the dilaton and the \RR\ scalar, and that the
dilaton and \RR\ scalar profile at $x$ is simply described by the
scalar propagator evaluated at $x$ and $x_0$. Under this assumption one
can calculate the interaction energy between the $D$-instanton and any
euclidean $D$-brane in field theory. As we want to demonstrate below,
this field theory calculation reproduces the contribution of the
lowest lying closed string states in the corresponding string cylinder
amplitude. The fact that these calculations agree gives support to the
identification of the $D$-instanton boundary state with the classical
solution above.

Let us consider a $D$-instanton (\ie\ a $(0,0)$-brane) at ${\bf y}$,
together with another euclidean brane at ${\bf 0}$. In order to get a
non-vanishing answer from the string calculation, the brane at
${\bf 0}$ will be taken to be, in turn, a
$\overline{(0,0)}$-brane, a $(2,0)$-brane and a
$(4,2)$-brane. Up to numerical constants, the contributions
to the corresponding cylinder diagrams from the zero modes
in the closed string channel
are given for $(0,0)-\overline{(0,0)}$ by (see (\ref{closeddef}))
\be
h_0({\bf y},{\bf 0})\, (4\pi m)^4\,
{q^{2m} \over  (1 - q^m)^4}
\sim h_0({\bf y},{\bf 0})\, (8\pi^2 \mu\, p^+)^4\,
\sin^{-4} (\pi\mu X^+)\,,
\label{partfnone}
\ee
while for the case of the $(0,0)$-brane at ${\bf y}$ and the
$(2,0)$-brane at the origin it is (see (\ref{DinstD1}))
\ba
j_0({\bf y}) &&\,
\sinh(\pi m)\, (4\pi m)^2  \, {q^{2m}\over (1-q^m)^3 (1+q^m)} \nn\\
&&\sim
j_0({\bf y}) \, \sinh(2\pi^2 \mu\, p^+) \,
(2\pi \mu\, p^+)^2\, \sin^{-3} (\pi\mu X^+) \cos^{-1}
(\pi \mu X^+)\,,
\label{partfntwo}
\ea
and for the the case of the $(0,0)$-brane at ${\bf y}$ and the
$(4,2)$-brane at the origin it is (see (\ref{DinstD5}))
\ba
j_0({\bf y}) &&\,
\sinh(\pi m)^{-1} (4\pi m)^2  \, {q^{2m}\over (1-q^m)^1(1+q^m)^3} \nn\\
&&\sim
j_0({\bf y}) \,\sinh(2\pi^2 \mu\, p^+)^{-1} \,
(2\pi \mu\, p^+)^2\,  \sin^{-1} (\pi\mu X^+) \cos^{-3} (\pi\mu X^+)\,.
\label{partfnthree}
\ea
Notice that the contribution of the $q^{2m}$ term to these expressions,
which can be isolated by letting $X^+ \to i\infty$, is proportional to
the charge $q_f$ (\ref{mixppinstf}) when this is evaluated in the same
limit.  This is to be expected since the cylinder diagram factorizes
into the product of the \RR\ charges of the ground states of the
two branes (one at the origin and one at $X^+$)  in this limit.

We will now show that the low energy limit of the right-hand sides of
these expressions coincide with the quantities obtained from the
exchange of a massless scalar field in this background. To see this
consider the situation in which there is the $(0,0)$-brane at
$x_1^+$, $x_1^-$ and ${\bf x_1}= {\bf y_1}$ in the presence of any
euclidean $(r,s)$-brane located at $x_2^+$, $x_2^-$ and
${\bf x_2}_t={\bf y_2}$ (where ${\bf x_2}_t$ is the transverse
position).  Then the force is proportional to the scalar propagator
between the $(0,0)$-brane  and the $(r,s)$-brane integrated
over the world-volume coordinates, ${\bf x_2}_l$, of the
$(r,s)$-brane. The propagator is simply equal to
\be
G(x_2;x_1) = (\Phi +i\epsilon)^{-4} \,,
\label{propdef}
\ee
where $\Phi$ is given in (\ref{origphi}). Next, take the Fourier
transform of the propagator with respect to $X^-=x_2^- -x_1^-$,
giving, for $X^+\equiv x_2^+ - x_1^+ >0$,
\be
p^+ \tilde G(X^+, {\bf x_1}, {\bf x_2}; p^+) = 2\pi (p^+)^4 \,
\left(\mu\over 2\sin (\pi \mu X^+)\right)^4
e^{{i\over 2}p^+\mu\,
\left({\bf x_2'}^2 \cot (\pi \mu X^+)+
{\bf x_1}^2 \tan(\pi  \mu X^+)\right)}  \,,
\label{fourierprop}
\ee
where the extra factor of $p^+$ has been included to conform to
the usual normalisation of the propagator in the light-cone gauge
and
\be
{\bf x_2'} = {\bf x_2} - {{\bf x_1}\over \cos\pi\mu X^+} \,.
\label{xpdef}
\ee
The full interaction is obtained by integrating this expression
over the  $p+1\equiv r+s$  ${\bf x_2}_l$ coordinates tangential
to the $(r,s)$-brane (letting ${\bf x_1} = {\bf y}$ and
setting ${\bf x_2}_t=0$ for simplicity),
\ba
& & p^+ \tilde G(X^+,  {\bf y}, {\bf x_2}_t=0; p^+) \nn\\
&&\quad = 2\pi \int  d^{p+1} {\bf x'_{2l}} \,
(p^+)^4 \, \left(\mu\over 2\sin (\pi \mu X^+)\right)^4 \nn\\
& & \qquad\qquad
e^{{i\over 2}p^+\mu \, \left({\bf x'_{2l}}^2 \cot(\pi  \mu X^+)
+ {\bf y_{l}}^2 \cot(\pi  \mu X^+)
+ {\bf y_{t}}^2 \tan(\pi  \mu X^+)\right)} \nn\\
&&\quad = 2\pi^{p+3\over 2}
(p^+)^{4-{p+1\over 2}}\,
\left({\mu\over 2}\right)^{4-{p+1\over 2}}\,
(\sin \pi \mu X^+)^{-4 + {p+1\over 2}}\,
(\cos \pi \mu X^+)^{-{p+1\over 2}}\,\nn\\
&&\qquad\qquad e^{{i\over 2}p^+\mu \,
\left( {\bf y_{l}}^2 \cot(\pi  \mu X^+) +
{\bf y_{t}}^2 \tan(\pi  \mu X^+)\right)} \,.
\label{finint}
\ea
This agrees with the expansion of the cylinder expressions
(\ref{partfnone}), (\ref{partfntwo}) and (\ref{partfnthree}) for
the cases $p=-1,$ $p=1$ and $p=5$ to leading order in $\alpha'$
(remembering that some of the prefactors contain powers of
$\sinh(2\pi^2\mu p^+) \simeq 2\pi^2 \mu p^+$).

\section{Discussion}
\label{discuss}

In this paper we have shown how to construct various
$D$-branes in the type IIB plane wave background that preserve half
the dynamical supersymmetries of the background. The instantonic
(or euclidean)
branes are characterised by the label $(r,s)$ ($r+s=p+1)$ which
defines the number of directions they occupy in the two $SO(4)$'s. In
\cite{bp,bgg} boundary states for class I $Dp$-branes with labels
$(0,2)$, $(2,0)$, $3,1)$, $(1,3)$, $(4,2)$ and $(2,4)$ were
constructed and the corresponding cylinder diagrams considered. In
those examples the open strings conserved half the kinematical
supersymmetries in addition to half the dynamical supersymmetries.
The cylinder diagrams joining pairs of these $D$-branes did not
vanish (when $p^+$ is fixed and non-zero).
Here we have generalized the earlier discussions
to include class II branes with the values $(0,0)$, $(0,4)$ and
$(4,0)$. These do not possess the kinematical supersymmetries.  On the
other hand, the open strings contain massless fermions which cause the
cylinder diagrams that link like pairs of branes to vanish. Given our
analysis and the independent considerations of \cite{hk} it seems that
these branes are the only class II branes that preserve dynamical
supersymmetries.

The interactions between pairs of $D$-branes associated with
cylindrical world-sheets were evaluated for a variety of cases.
In each case the cylinder was evaluated in the closed-string
channel as an overlap of two boundary states, and also as a trace
over the states of an open string joining the two $D$-branes.   A
rather nontrivial behaviour of these expressions under the $S$ modular
transformation was needed in order for the closed-string and
open-string calculations to agree.

As already described in \cite{bgg} our discussion generalises
directly to the case of lorentzian signature $D$-branes in which the
light-cone coordinates, $x^\pm$, are tangential to the
world-volume and are of the form $(+,-;r,s)$.
In fact, the open string analysis we have presented in
this paper can be directly applied to time-like branes if we consider
the usual open string light cone gauge, and replace $\hm$ by
$m$. Apart from the supersymmetric $D$-branes that had been considered
in \cite{dp}, we have also constructed supersymmetric (class II)
branes of types $(+,-;0,0)$, $(+,-;4,0)$ and $(+,-;0,4)$. The latter
two carry non-trivial Born-Infeld flux that is induced by the
background \RR\ flux. As a consequence, the corresponding
`anti-branes' break supersymmetry.  As remarked earlier, it should be
simple to generalize our discussion to include oblique branes \cite{hk}
which are oriented in directions that link the first and second
$SO(4)$'s and therefore
cannot be represented in the form $(+,-;r,s)$.
Furthermore,
the light-cone gauges we have used are not appropriate for describing
$D$-branes in which $x^-$ is tangential to the brane with $x^+$ being
transverse (or the converse), which includes the case of `null branes'.

It would be interesting to understand the properties of these branes
from a more geometrical viewpoint.  For example, it is clear that the
$(+,-;0,0)$ D-string arises as a Penrose limit of a
non-supersymmetric  string wrapping the equator of the five-sphere in
$AdS^5\times S^5$ \cite{bpz,st}.  Likewise, the $(+,-;0,4)$-brane
comes from the baryonic vertex of $AdS^5\times S^5$, which is a
supersymmetric $D5$-brane wrapping the five-sphere. On the other hand,
the $(+,-;4,0)$-brane originates from a non-supersymmetric $D5$-brane
in  $AdS^5\times S^5$ one of whose axis wraps an equator of the
five-sphere while the other axes are aligned with directions in
$AdS^5$.

The connection of the instantonic branes we have constructed with
instantonic branes in $AdS^5\times S^5$ is more obscure.
Understanding this could lead to an understanding of the relation between
the $D$-instanton and instanton effects in the dual Yang--Mills
field theory. It would also be interesting to understand the effect of
the $D$-instanton on the plane-wave dynamics. Finally, one should be
able to analyse the $D$-instanton contributions by considering the
effects of the $R^4$ and related terms in the effective low energy IIB
action \cite{gg2} in this background.
\bigskip

\noindent {\bf Note added:} while this paper was in the final stages
of preparation a preprint by Skenderis and Taylor appeared \cite{st2}
which contains results that overlap with our section 3.1.1 and
appendix C.

\section*{Acknowledgements}

We are grateful to Juan Maldacena for useful conversations, and to
Pascal Bain, Terry Gannon, Eric Gimon, Leo Pando Zayas, Aninda Sinha
and Nemani Suryanarayana for discussions.

MRG is grateful to the Royal Society for a University Research
Fellowship. He thanks the Institute for Advanced Study for hospitality
while this paper was being completed; his research there was supported
by a grant in aid from the Funds for Natural Sciences. We also
acknowledge  partial support from the PPARC Special Programme Grant
`String Theory and Realistic Field Theory', PPA/G/S/1998/0061 and the
EEC contract HPRN-2000-00122.

\appendix

\section{Notation and definitions}
\label{notate}

We shall be using the same conventions as in \cite{bgg} (see also
\cite{mt,bp}). The light-cone lagrangian in the plane-wave background
describes eight massive free scalar and eight massive free fermion
fields,
\be
{\cal L} = {1\over 4\pi} \left( \partial_+ x^I \partial_- x^I
- m^2 (x^I)^2  \right)
+ {i \over 2\pi}\left(S^a\partial_+ S^a + \tilde S^a \partial_- \tilde S^a
- 2m\, S^a\, \Pi_{ab} \, \tilde S^b \right)\, ,
\label{lcact}
\ee
where $S^a$ and $\tilde S^a$ are $SO(8)$ spinors of the same chirality
and $\Pi = \gamma^1 \gamma^2\gamma^3 \gamma^4$. The mass parameter
$m$ is defined by $m=2 \pi p^+ \mu$. The $8\times 8$ matrices,
$\gamma^I_{a\dot b}$ and $\gamma^I_{\dot ab}$, are the
off-diagonal  blocks of the $16\times 16$ $SO(8)$ $\gamma$-matrices
and couple $SO(8)$ spinors of opposite chirality.  The presence of
$\Pi$ in the fermionic sector of the lagrangian breaks the symmetry
from $SO(8)$ to $SO(4)\times SO(4)$.

For the bosonic degrees of freedom, the general solution to the
equations of motions takes the form
\ba
x^I(\sigma,\tau) & = & \cos(m\tau) x_0^I + {1\over m} \sin(m\tau)
p_0^I + i \sum_{n\ne 0} {1\over \omega_n}
\left( e^{-i(\omega_n\tau - n \sigma)} \alpha^I_n +
e^{-i(\omega_n\tau + n \sigma)} \tilde\alpha^I_n \right)\,, \nn \\
{\cal P}^I(\sigma,\tau) & = & \cos(m\tau) p_0^I - m \sin(m\tau)
x_0^I + \sum_{n\ne 0}
\left( e^{-i(\omega_n\tau - n \sigma)} \alpha^I_n +
e^{-i(\omega_n\tau + n \sigma)} \tilde\alpha^I_n \right) \,,
\ea
where ${\cal P}^I$ is the canonical momentum associated to $x^I$,
${\cal P}^I= \dot{x}^I$ \cite{met}. The non-zero modes $\alpha^I_k$
and $\tilde{\alpha}^I_k$ satisfy the commutation relations
\be
[\alpha^I_k,\alpha^J_l]  = \omega_k \, \delta^{IJ}\, \delta_{k,-l}
\,, \qquad
[\alpha^I_k,\tilde\alpha^J_l]  = 0 \, , \qquad
[\tilde\alpha^I_k,\tilde\alpha^J_l] = \omega_k\, \delta^{IJ}\,
\delta_{k,-l} \,,
\label{boscom}
\ee
where
\be
\omega_k = {\rm sign}(k)\, \sqrt{k^2 + \m^2}\qquad |k|>0\, .
\label{omegadef}
\ee
On the other hand, the centre of mass position $x^I_0$ and the
generalised momentum $p^I_0$ commute with the non-zero modes above,
and satisfy
\be
[p_0^I,x_0^J] = - i \delta^{IJ} \,.
\label{boszero}
\ee
It is convenient to introduce the creation and annihilation operators
\be
a^I_0 = {1\over \sqrt{2\m}} \bigl(p_0^I + i \m x_0^I\bigr) \,, \qquad
\bar{a}^I_0 = {1\over \sqrt{2\m}} \bigl(p_0^I - i \m x_0^I\bigr) \,,
\label{creani}
\ee
in terms of which (\ref{boszero}) is then simply
\be
[\bar{a}_0^I,a_0^J] = \delta^{IJ}\,.
\label{creanii}
\ee
The fermionic fields can be similarly expanded in terms of modes
\begin{eqnarray}\label{modeexpansionclosed}
S(\sigma,\tau) & = & S_0 \cos(m\tau) + \Pi \tilde{S}_0 \sin(m\tau)
\nn\\
& & \qquad + \sum_{n\ne 0} c_n \left[
S_n\,  e^{-i(\omega_n\tau - n \sigma)}
+   {i\over m} (\omega_n-n) \Pi \tilde{S}_n
        e^{-i(\omega_n\tau + n \sigma)} \right] \,, \\
\tilde{S}(\sigma,\tau) & = &
-\Pi S_0 \sin(m\tau) + \tilde{S}_0 \cos(m\tau)  \nn\\
& & \qquad + \sum_{n\ne 0} c_n \left[
    \tilde{S}_n e^{-i(\omega_n\tau + n \sigma)}
- {i\over m} (\omega_n-n) \Pi S_n e^{-i(\omega_n\tau - n \sigma)}
\right]\,,
\end{eqnarray}
where $c_n$ is defined  by
\be
c_n ={m \over \sqrt{2\omega_n (\omega_n -n)}}\, .
\label{cndef}
\ee
The modes $S^a_k$ and $\tilde{S}^a_k$, where $a$ is a
spinor index of $SO(8)$ and $k\in\Zop$, satisfy the anti-commutation
relations 
\be
\{ S^a_k,S^b_l\} = \delta^{ab}\delta_{k,-l} \,,  \qquad
\{ S^a_k,\tilde{S}^b_l\}  = 0 \,, \qquad
\{\tilde{S}^a_k,\tilde{S}^b_l\}  = \delta^{ab}\delta_{k,-l} \,.
\label{fermcom}
\ee
It is convenient to introduce the zero-mode combinations,
\be
\theta^a_0 = {1\over \sqrt{2}}(S^a_0 + i \tilde{S}^a_0)
\,, \qquad \bar\theta^a_0 = {1\over \sqrt{2}}(S^a_0 - i \tilde{S}^a_0)
\,,
\label{thetadef}
\ee
and further
\ba
\theta_R & =& \half (1 + \Pi) \theta_0 \,,  \qquad
\bar\theta_R = \half (1 + \Pi) \bar\theta_0 \,, \nn\\
\theta_L & =& \half (1 - \Pi) \theta_0 \,, \qquad
\bar\theta_L = \half (1 - \Pi) \bar\theta_0 \,.
\label{thetarl}
\ea
The dynamical supercharges of the closed string theory are given by
\cite{mt,bp}\footnote{We are adopting a slightly  different
normalisation for the non-zero mode contributions.}
\begin{eqnarray}
\sqrt{ 2 p^+}\, Q_{\dot a} & = & {1\over 2\pi} \int_0^{2\pi} d\sigma 
\left[ {\cal P}^I \gamma^I S - {x'}^I \gamma^I S 
- m x^I \gamma^I \Pi \tilde{S}  \right] \nn\\
\sqrt{ 2 p^+}\, \tilde{Q}_{\dot a} & = & 
{1\over 2\pi} \int_0^{2\pi} d\sigma 
\left[ {\cal P}^I \gamma^I \tilde{S} + {x'}^I \gamma^I \tilde{S} 
+ m x^I \gamma^I \Pi S  \right] \,,
\end{eqnarray}
which in terms of modes is 
\begin{eqnarray}
\label{q-}
\sqrt{ 2 p^+}\, Q_{\dot a}
&=&  p_0^I \gamma^I_{\dot a b} S_0^b
   - m x_0^I \left(\gamma^I\Pi\right)_{\dot a b}
  {\tilde S}_0^b
\nonumber\\
&+& \sum_{n=1}^\infty
\left( c_n \gamma^{I}_{\dot a b}
       (\alpha_{-n}^I S_n^b + \alpha_n^I S^{b}_{-n} )
+
 \frac{{\rm i} m}{2\omega_n c_n }
 \left(\gamma^I\Pi\right)_{\dot a b}
       ({\tilde \alpha}_{-n}^I {\tilde S}_n^b
        -{\tilde \alpha}_n^I {\tilde S}^b_{-n} )
\right)\,,
\nonumber\\
\\
\label{qt-}
\sqrt{ 2 p^+}\,  {\tilde Q}_{\dot a}
&=& p_0^I \gamma^{I}_{\dot a b} {\tilde S}_0^b
  + m x_0^I\left(\gamma^I\Pi\right)_{\dot a b}{ S}_0^b
\nonumber\\
&+& \sum_{n=1}^\infty
\left(( c_n  \gamma^{I}_{\dot a b}
       ({\tilde \alpha}_{-n}^I {\tilde S}_n^b
        +{\tilde \alpha}_n^I {\tilde S}^b_{-n} )
-
 \frac{{\rm i} m}{2 \omega_n c_n }
 \left(\gamma^I\Pi\right)_{\dot a b}
       (\alpha_{-n}^I S_n^b - \alpha_n^I S^b_{-n})
\right)\,.
\nonumber\\
\end{eqnarray}
In order to describe the anti-commutation relations of the dynamical
supercharges it is useful to introduce
$Q^\pm_{\dot a}={1\over\sqrt{2}}(Q_{\dot a} \pm i \tilde{Q}_{\dot a})$.
(Note that in contradistinction to (\ref{superp}) and (\ref{superm}),
for example, the index $\pm$ here does not indicate the eigenvalue
with respect to the action of $\Pi$.) Then the anti-commutation
relations are \cite{mt}
$\{ Q^\pm_{\dot a}, Q^\pm_{\dot b} \} = 0$, as well as
\be
\{ Q^+_{\dot a}, Q^-_{\dot b} \}  = 2\, \delta_{\dot a\dot b}\, H +
m\, (\gamma^{ij} \,\Pi)_{\dot a \dot b} \, J^{ij}
+ m \, (\gamma^{i'j'}\, \Pi)_{\dot a \dot b} \, J^{i'j'}  \,,
\ee
where $J^{ij}$ are the rotation generators (see \cite{mt}) while $H$
is the light-cone hamiltonian $H$ for the closed string in the plane-wave
background
\ba
2 \, p^+ H &=& \m \left(a^I_0\, \bar{a}^I_0
+ i\, S^a_0 \,\Pi_{ab}\, \tilde{S}^b_0 +4\right)
+ \sum_{k=1}^{\infty} \left[ \alpha^I_{-k}\alpha^I_k +
\tilde\alpha^I_{-k} \tilde\alpha^I_k
+ \omega_k \left(S^a_{-k} S^a_{k} + \tilde{S}^a_{-k} \tilde{S}^a_{k}
\right) \right]\nn\\
 &=& \m \left(a^I_0\, \bar{a}^I_0 + \theta_L^a \,\bar\theta_L^a +
\bar\theta_R^a \,\theta_R^a \right)
+  \sum_{k=1}^{\infty} \left[
\alpha^I_{-k}\alpha^I_k +
\tilde\alpha^I_{-k} \tilde\alpha^I_k
+ \omega_k \left(S^a_{-k} S^a_{k} + \tilde{S}^a_{-k} \tilde{S}^a_{k}
\right) \right]
\,.\nn\\
\label{lcham}
\ea
In the limit $m\equiv 2\pi p^+\mu \to 0$ this reduces to the usual
light-cone gauge hamiltonian in a flat background \cite{gs1}. The
normal ordering has been chosen in (\ref{lcham}) with the
understanding that $\theta_L^a$ and $\bar\theta_R^a$ are creation
operators while $\bar\theta_L^a$ and $\theta_R^a$ are annihilation
operators.

As is familiar from flat space, the space of states is described by a
tensor product of the space generated by the bosonic modes and that
generated by the fermionic modes. The ground state of the bosonic
space, $|0\rangle_b$, is annihilated by the modes $\bar{a}^I_0$ as
well as $\alpha^I_k$ and $\tilde{\alpha}^I_k$ with $k>0$ and is
non-degenerate since each of the `zero modes' $a^I_0$ raises the
energy by $\m$.  Likewise, the non-degenerate ground state in the
space spanned by the fermionic operators, $|0\rangle_f$, is the state
annihilated by $\bar\theta_L^a$ and $\theta_R^a$, while the creation
operators $\theta_L^a$ and $\bar\theta_R^a$ raise the energy by $m$.

Finally, the kinematical supercharges, $Q_a\equiv S^a_0$ and
$\tilde{Q}_a\equiv \tilde{S}^a_0$ do
not commute with $H$, but rather satisfy
\be\label{hamilcom}
[H, Q_a] = - {i m \over 2 p^+} \,\Pi_{ab}\, \tilde{Q}_b \,,\qquad
[H, \tilde{Q}_a] =  {i m \over 2 p^+}\, \Pi_{ab}\, Q_b \,.
\ee

\section{Definition of $f_i$, $g_i$, $\hat g_i$, $g_{i,\pm}$,
$\hat g_{i,\pm}$}
\label{appa}

The expressions for the cylinder diagrams with boundaries on pairs
of class I branes discussed in \cite{bgg} are defined by
\ba
f_1^{(\m)}(q) & =& q^{-\Delta_\m} (1-q^\m)^{{1\over 2}}
\prod_{n=1}^{\infty} \left(1 - q^{\sqrt{\m^2+n^2}}\right) \,,
\label{fdef}\\
f_2^{(\m)}(q) & = &q^{- \Delta_\m} (1+q^\m)^{{1\over 2}}
\prod_{n=1}^{\infty} \left(1 + q^{\sqrt{\m^2+n^2}}\right) \,,
\label{f2def}\\
f_3^{(\m)}(q) & =& q^{-\Delta^\prime_\m}
\prod_{n=1}^{\infty} \left(1 + q^{\sqrt{\m^2+(n-1/2)^2}}\right) \,,
\label{f3def}\\
f_4^{(\m)}(q) & =& q^{-\Delta^\prime_\m}
\prod_{n=1}^{\infty} \left(1 - q^{\sqrt{\m^2+(n-1/2)^2}}\right) \,,
\label{f4def}
\ea
where $\Delta_\m$ and $\Delta^\prime_\m$ are given as
\ba
\Delta_\m & =& -{1\over (2\pi)^2} \sum_{p=1}^{\infty}
\int_0^\infty ds \, e^{-p^2 s} e^{-\pi^2 \m^2 / s} \,, \cr
\Delta^\prime_\m & =& -{1\over (2\pi)^2} \sum_{p=1}^{\infty}
(-1)^p \int_0^\infty ds \, e^{-p^2 s} e^{-\pi^2 \m^2 / s}\,.
\label{Deltadef}
\ea
These functions satisfy the conditions
\be
f_1^{(\m)}(q) = f_1^{(\hm)}(\tq), \qquad f_2^{(\m)}(q) =
f_4^{(\hm)}(\tq), \qquad f_3^{(\m)}(q) = f_3^{(\hm)}(\tq).
\label{smods}
\ee

In this paper various other cylinder diagrams arise that are
expressed in terms of generalizations of the above functions.  The
complete list of these functions is as follows,
\be
g_1^{(m)}(t) = 4\pi \,i\, {m}\, q^{-2\Delta_m}\, q^{m/2}\,
\prod_{n=1}^{\infty}
\left( 1 - \left(\omega_n+m \over \omega_n-m\right)
q^{\omega_n} \right)
\left( 1 - \left(\omega_n-m \over \omega_n+m\right)
q^{\omega_n} \right)\,,
\label{g1deff}
\ee
\be
g_2^{(m)}(t) = 4\pi {m}\, q^{-2\Delta_m}\, q^{m/2}\,
\prod_{n=1}^{\infty}
\left( 1 + \left(\omega_n+m \over \omega_n-m\right)
q^{\omega_n} \right)
\left( 1 + \left(\omega_n-m \over \omega_n+m\right)
q^{\omega_n} \right)\,,
\label{g2def}
\ee
\be\label{g3def}
g_3^{(m)}(t) = 2\, q^{-2\Delta'_m}\,
\prod_{n=1}^{\infty}
\left( 1 + \left(\omega_{n-1/2}+m \over \omega_{n-1/2}-m\right)
q^{\omega_{n-1/2}} \right)
\left( 1 + \left(\omega_{n-1/2}-m \over \omega_{n-1/2}+m\right)
q^{\omega_{n-1/2}} \right)\,,
\ee
\be\label{g4deff}
g_4^{(m)}(t) = 2\, q^{-2\Delta'_m}\,
\prod_{n=1}^{\infty}
\left( 1 - \left(\omega_{n-1/2}+m \over \omega_{n-1/2}-m\right)
q^{\omega_{n-1/2}} \right)
\left( 1 - \left(\omega_{n-1/2}-m \over \omega_{n-1/2}+m\right)
q^{\omega_{n-1/2}} \right)\,,
\ee
\be\label{g1hatdeff}
\hat{g}_1^{(\hm)}(\ttt) = \tq^{-\widetilde{\Delta}_{\hm}}
\prod_{l\in{\cal M}_+} \left(1-\tq^{\,\hom_l}\right)^{1\over 2}
\prod_{l\in{\cal M}_-} \left(1-\tq^{\,\hom_l}\right)^{1\over 2} \,,
\ee
\be\label{g2hatdef}
\hat{g}_2^{(\hm)}(\ttt) = \tq^{-\widetilde{\Delta}_\hm}\,
\prod_{l\in{\cal M}_+}
\left(1+\tilde{q}^{\,\widehat\omega_l}\right)^{1\over 2}
\prod_{l\in{\cal M}_-}
\left(1+\tilde{q}^{\,\widehat\omega_l}\right)^{1\over 2} \,,
\ee
\be\label{g3hatdef}
\hat{g}_3^{(\hm)}(\ttt) = \tq^{-\widehat{\Delta}_\hm}\,
\prod_{l\in{\cal P}_+}
\left(1+\tilde{q}^{\,\widehat\omega_l}\right)^{1\over 2}
\prod_{l\in{\cal P}_-}
\left(1+\tilde{q}^{\,\widehat\omega_l}\right)^{1\over 2} \,,
\ee
\be
\hat{g}^{(\hm)}_4(\tilde t) = \tilde{q}^{-\widehat{\Delta}_{\hm}}
\prod_{l\in{\cal P}_+}
\left(1-\tilde{q}^{\,\widehat\omega_l}\right)^{1\over 2}
\prod_{l\in{\cal P}_-}
\left(1-\tilde{q}^{\,\widehat\omega_l}\right)^{1\over 2} \,.
\label{g4hatdeff}
\ee
In these expressions
$\hom_k={\rm sign}(k)\, \sqrt{k^2 + \hm^2}$ and
\ba
\widetilde \Delta_{\hm} &=&  - {1\over (2\pi)^2}
\sum_{p=1}^{\infty} \sum_{r=0}^{\infty}
c_r^p\, \hm\, {\partial^r \over (\partial  \hm^2)^r}
{1\over \hm} \int_0^\infty ds
\left({-s\over \pi^2 }\right)^r
e^{-p^2 s -\pi^2 \hm^2 /s} \,,\nn\\
\widehat{\Delta}_{\hm} &=& - {1\over (2\pi)^2}
\sum_{p=1}^{\infty} (-1)^p \sum_{r=0}^{\infty}
c_r^p\, \hm\, {\partial^r \over (\partial  \hm^2)^r}
{1\over \hm} \int_0^\infty ds
\left({-s\over \pi^2 }\right)^r
e^{-p^2 s -\pi^2 \hm^2 /s} \,,
\label{deltsdefs}
\ea
where $c_r^p$ are the Taylor coefficients of the functions
\be\label{cdef}
\left({x+1\over x-1}\right)^p + \left({x-1\over x+1}\right)^p =
\sum_{r=0}^\infty c_r^p\, x^{2r} \,.
\ee
Furthermore, the sets ${\cal P}_\pm$ and ${\cal M}_{\pm}$ are defined
by
\ba
l\in {\cal P}_+ & \qquad \hbox{if} \quad  &
{l+i\hm \over l-i\hm} + e^{2\pi i l} =0\,,\\
l\in {\cal P}_- & \qquad \hbox{if} \quad  &
{l-i\hm \over l+i\hm} + e^{2\pi i l} =0 \,, \\
l\in {\cal M}_+ & \qquad \hbox{if} \quad &
{l + i\hm \over l - i\hm} - e^{2\pi i l} =0\,,
\label{calmplusdeft} \\
l\in {\cal M}_- & \qquad \hbox{if} \quad &
{l - i\hm \over l + i\hm} - e^{2\pi i l} =0\,.
\label{calpplusdeft}
\ea
We shall show in detail in appendix~\ref{modprops} that
\be
g_2^{(m)}(t)=\hat{g}_4^{(\hm)}(\ttt)\, .
\label{g2gh4}
\ee
The contour integral method that we use is easily extended to show the
other modular properties,
\be
g_1^{(m)}(t)=\hat{g}_1^{(\hm)}(\ttt),\qquad
g_4^{(m)}(t)=\hat{g}_2^{(\hm)}(\ttt), \qquad
g_3^{(m)}(t)=\hat{g}_3^{(\hm)}(\ttt) \,.
\label{alrest}
\ee
For $m\rightarrow 0$, these relations reduce to the standard modular
formulae relations between $f_1$, $f_2$, $f_3$ and $f_4$. Let us, for
example, consider the case of the first relation in
(\ref{alrest}). For $\hm$  small, there are two  solutions with
$l\in {\cal M}_+$ close to zero at $l=\pm \sqrt{\hm/\pi}$, and two more
solutions with  $l\in {\cal M}_-$ close to zero at
$l=\pm i \sqrt{\hm/\pi}$.  Consequently, for small $\hm$,
$\hat{g}_1^{(\hm)}(\ttt) \sim 4\,\pi\,\hm\, i\, \ttt^2\,\eta^2(\ttt)$,
while $g_1^{(m)}(t)\sim 4\,\pi\, m\, i\, \eta^2(t)$, and we
recover the standard modular transformation formula of the
$\eta$-function.

\subsection{The factorized $g$-functions}
\label{factorg}

The functions $g_i^{(m)}(t)$ and $\hat g_i^{(\hm)}(t)$ are naturally
expressed in factorized form
\be
g_i^{(m)}(t) = g_{i,-}^{(m)}(t)\; g_{i,+}^{(m)}(t)\,, \qquad
\hat g_i^{(\hm)}(t) =\hat{g}_{i,-}^{(\hm)}(\tilde t) \;
\hat{g}_{i,+}^{(\hm)}(\tilde t)\,.
\label{facone}
\ee
The individual factors are defined in such a manner that they have
simple transformation properties under the $S$ modular
transformation.

We will omit the derivation of these functions and simply
state the explicit relations for the cases that are relevant for
us. For the closed string functions define
\ba
g_{1,+}^{(m)}(t) & = & 4 \, \pi \, i\, m\,  e^{{\cal D}_m} \, q^{m/2}\,
q^{-\Delta_m}\,
\prod_{n=1}^{\infty}
\left( 1 - \left(\omega_n+m \over \omega_n-m\right)
q^{\omega_n} \right)\,,\\
g_{1,-}^{(m)}(t) & = & e^{{-\cal D}_m} \, q^{-\Delta_m}\,
\prod_{n=1}^{\infty}
\left( 1 - \left(\omega_n-m \over \omega_n+m\right)
q^{\omega_n} \right)\,, \\
g_{2,+}^{(m)}(t) & = & 4 \, \pi \, m\,  e^{{\cal D}_m} \, q^{m/2}\,
q^{-\Delta_m}\,
\prod_{n=1}^{\infty}
\left( 1 + \left(\omega_n+m \over \omega_n-m\right)
q^{\omega_n} \right)\,,\\
g_{2,-}^{(m)}(t) & = & e^{{-\cal D}_m} \, q^{-\Delta_m}\,
\prod_{n=1}^{\infty}
\left( 1 + \left(\omega_n-m \over \omega_n+m\right)
q^{\omega_n} \right)\,,
\ea
where ${\cal D}_m$ is defined by
\be
\label{Ddef}
{\cal D}_m = {1\over 2 \sqrt{\pi}}
\sum_{\hn=1}^{\infty}
\int_0^\infty {d\ts \over \ts^{1/2}} \,e^{-\hn^2 \ts}\,
\hbox{Erf}(\pi m/\ts^{1/2}) \,,
\ee
with Erf$(x)$ the error function
\be\label{erf}
\hbox{Erf}(x) = 1 - {2\over\sqrt{\pi}} \int_0^x  du \,
e^{-u^2}\,.
\ee
Under the $S$-modular transformation, these functions transform as
\be
g_{1,\pm}^{(m)}(t) = \hat{g}_{1,\mp}^{(\hm)}(\tilde t)\,, \qquad
g_{2,\pm}^{(m)}(t) = \hat{g}_{4,\mp}^{(\hm)}(\tilde t)\,,
\ee
where the functions on the right hand side are defined by
\ba
\hat{g}^{(\hm)}_{1,\pm}(\tilde t) & = &
\tilde{q}^{-\widetilde{\Delta}_{\hm,\pm}}
\prod_{l\in{\cal M}_\pm}
\left(1-\tilde{q}^{|\widehat\omega_l|}\right)^{1\over 2} \,,
\label{hatg1+def}\\
\hat{g}^{(\hm)}_{4,\pm}(\tilde t) & = &
\tilde{q}^{-\widehat{\Delta}_{\hm,\pm}}
\prod_{l\in{\cal P}_\pm}
\left(1-\tilde{q}^{|\widehat\omega_l|}\right)^{1\over 2} \,.
\label{hatg4+def}
\ea
Here
\ba
\widetilde\Delta_{\hm,\epsilon} & = & - {1\over (2\pi)^2}
\sum_{p=1}^{\infty}
{1\over p}  \sum_{r=0}^\infty d^p_r \,
\left( {\epsilon \over 2\pi \hm} \right)^r
{\partial^r \over (\partial p)^r} \, p
\int_0^\infty ds\,
e^{-p^2 s} e^{-\pi^2 \hm^2/s}\,, \\
\widehat\Delta_{\hm,\epsilon} & = & - {1\over (2\pi)^2}
\sum_{p=1}^{\infty}
{(-1)^p\over p}  \sum_{r=0}^\infty d^p_r \,
\left( {\epsilon \over 2\pi \hm} \right)^r
{\partial^r \over (\partial p)^r} \, p
\int_0^\infty ds\,
e^{-p^2 s} e^{-\pi^2 \hm^2/s}\,,
\ea
where $d^p_r$ are the Taylor coefficients of the functions
\be
\left( { x+1 \over x-1}\right)^p = \sum_{r=0}^{\infty} d_r^p\, x^r \,.
\ee

\section{Supersymmetry of the open string}

In this appendix the anti-commutation relations of the open-string
modes will be determined.  We will also show that the supercharge
(\ref{super}) is indeed time-independent.

\subsection{Anti-commutation relations of the modes}
\label{susyanti}

The anti-commutation relations of the modes $S_n$ are fixed
by the requirement that the fields $S$ and $\tilde{S}$ satisfy the
usual equal time anti-commutation relations
\ba
\left\{S^a (\sigma,\tau),  S^b (\sigma',\tau)
\right\} &= & 2\,\pi\, \delta^{ab}\, \delta(\sigma -\sigma')\,,\nn\\
\left\{S^a (\sigma,\tau),  \tilde{S}^b (\sigma',\tau)
\right\} &= & 0 \,, \label{anticomms} \\
\left\{\tilde{S}^a (\sigma,\tau),  \tilde{S}^b (\sigma',\tau)
\right\} &= & 2\,\pi\, \delta^{ab}\, \delta(\sigma -\sigma')\,,\nn
\ea
where $0<\sigma,\sigma'<\pi$. These are equivalent to
\ba
\left\{S^a (\sigma,\tau)\pm \tilde{S}^a(\sigma,\tau),
S^b (\sigma',\tau) \pm \tilde{S}^b(\sigma',\tau)
\right\} &= & 4\,\pi\, \delta^{ab}\, \delta(\sigma -\sigma')\,,
\label{anticomms1} \\
\left\{S^a (\sigma,\tau)+ \tilde{S}^a(\sigma,\tau),
S^b (\sigma',\tau) - \tilde{S}^b(\sigma',\tau)
\right\} &= & 0 \,.\label{anticomms2}
\ea
Given the mode expansions (\ref{sexpans}) it is easy to see that the
relation (\ref{anticomms2}) as well as the relation in
(\ref{anticomms1}) involving $S-\tilde{S}$ are satisfied provided that
\be\label{anti1}
\{ S^a_n, S^b_m \} = \delta^{a,b}\, \delta_{n,-m} \,,\qquad
\hbox{if $n\ne 0$ or $m\ne 0$.}
\ee
Here one uses the standard identity
\be
\sum_{n=-\infty}^\infty e^{inx} = 2\,\pi\, \sum_{r\in\Zop}
\delta (x + 2\pi r) \,,
\label{deltads}
\ee
as well as the fact that for $0< \sigma,\sigma' <\pi$,
$\delta(\sigma+\sigma' + 2 \pi r)=0$ for all $r\in\Zop$.

Given (\ref{anti1}) as well as the mode expansion of $S+\tilde{S}$
in (\ref{sexpans}) it is straightforward to determine the
contribution of the non-zero modes to
$\{(S+\tilde{S})(\sigma,\tau),(S+\tilde{S})(\sigma',\tau)\}$, which is
\be
4\,\pi\, \delta^{ab} \delta(\sigma-\sigma') +
2 \,\delta^{ab}\, \sum_{n\in\Zop} {n^2 - \hm^2 \over n^2+\hm^2}
e^{i n (\sigma+\sigma')}
+ 2 \, \Pi^{ab} \sum_{n\in\Zop} {-2n\hm i \over n^2+\hm^2}
e^{i n (\sigma+\sigma')} \,.
\ee
Note that the two infinite sums now run over all integers; this is
immaterial for the second sum, but in the first sum the contribution
from $n=0$ cancels the $n=0$ contribution that is necessary in order
to produce the $\delta$-function of the first term via
(\ref{deltads}).

The two infinite sums can be evaluated by replacing the sum by a
contour integral; for the case of the first sum the relevant contour
integral is
\be
\sum_{n\in\Zop} {n^2 - \hm^2 \over n^2+\hm^2}
e^{i n (\sigma+\sigma')}
= - \oint_C {d\nu\over 1-e^{2\pi i\nu}}
{\nu^2-\hm^2 \over \nu^2+\hm^2}  e^{i\nu(\sigma+\sigma')} \,,
\ee
where the contour consists of two lines passing infinitesimally above
and below the real axis. Since $0<\sigma+\sigma'$ the upper contour
can be closed at infinity in the upper half plane and picks up the
contribution from the pole at $\nu=i\hm$ to give
\be\label{I1}
I_1 =  - 2\, \pi {\hm \over 1-e^{-2\pi\hm}} e^{-\hm (\sigma+\sigma')}
\,.
\ee
Similarly, since $\sigma+\sigma'<2\pi$, the lower contour can be
closed at infinity in the lower half plane, and gives rise to
\be\label{I2}
I_2 =  2\, \pi {\hm \over 1-e^{2\pi\hm}} e^{\hm (\sigma+\sigma')}
\ee
from the pole at $-i\hm$. By an analogous calculation one finds
\be\label{I3}
\sum_{n\in\Zop} {-2n\hm i \over n^2+\hm^2}
e^{i n (\sigma+\sigma')} =
2 \,\pi \, {\hm \over 1-e^{-2\pi\hm}} e^{-\hm (\sigma+\sigma')}
+ 2  \,\pi \, {\hm \over 1-e^{2\pi\hm}} e^{\hm (\sigma+\sigma')}\,.
\ee
In order to reproduce (\ref{anticomms1}), the sum of (\ref{I1}),
(\ref{I2}) and (\ref{I3}) must be cancelled by the contribution from
the zero modes to this anti-commutator. This is precisely the case
provided that the anti-commutator of the zero modes is given as in
(\ref{fermioncom2}).

\subsection{Dynamical supercharge}
\label{dynsusy}

The fact that the dynamical supercharge is time independent simply
follows by substituting (\ref{sexpans}) and (\ref{bosonmode}) into
(\ref{super}). In order to see this let us first consider the terms
involving bosonic or fermionic zero modes. The contribution that is
proportional to  $y_1^I \gamma^I S_n$ with $n\ne 0$
is equal to  (up to the
irrelevant prefactor of $\hm / 2 \sqrt{X^+}$)
\ba
y_1^I &\gamma^I &\sum_{n\ne 0} e^{-i\hom_n\tau} \left[
\int_0^\pi d\sigma \sinh(\hm\sigma)
\left( \cos(n\sigma) + {\hm \over n} \sin(n\sigma) \Pi\right)
\left[ {n\over \hom_n} + i {n\,(\hom_n-n) \over \hm\,\hom_n}  \Pi
\right] S_n \right. \nn\\
& + & \left.
\int_0^\pi d\sigma \cosh(\hm\sigma)
\left( \cos(n\sigma)\Pi + {\hm \over n} \sin(n\sigma) \right)
\left[ {n\over \hom_n} + i {n\,(\hom_n-n) \over \hm\,\hom_n} \, \Pi
\right] S_n \right]\,. \label{zerodanger}
\ea
Using the identities
\ba
\int_0^\pi d\sigma \cosh(\hm\sigma)\, \cos(n\sigma)  & = &
{\hm (-1)^n \over \hm^2+n^2}\, \sinh(\hm\pi) \,, \nn\\
\int_0^\pi d\sigma \sinh(\hm\sigma)\, \sin(n\sigma)  & = &
- {n (-1)^n \over \hm^2+n^2}\, \sinh(\hm\pi) \,, \nn\\
\int_0^\pi d\sigma \cosh(\hm\sigma)\, \sin(n\sigma)  & = &
- {n \over \hm^2+n^2}\, \left[(-1)^n \cosh(\hm\pi) - 1 \right] \,,
\nn\\
\int_0^\pi d\sigma \sinh(\hm\sigma)\, \cos(n\sigma)  & = &
{\hm \over \hm^2+n^2}\,\left[(-1)^n \cosh(\hm\pi) - 1 \right]
\label{coshsin}
\ea
it is easy to see that (\ref{zerodanger}) vanishes. The analysis is
identical for the term proportional to
$(y_2^I - \cosh(\hm\pi) y_1^I) \gamma^I S_n$ with $n\ne 0$.

The analysis is similar for the terms
proportional to $\alpha^I_l \gamma^I S_0$ with
$l\ne 0$, \ie\
\ba
{2\over\sqrt{X^+}} \sum_{l\ne 0} e^{-i\hom_l\tau}&& \left[
{l\over \hom_l} \alpha^I_l \gamma^I
\left( \cos(l\sigma) S_0 \cosh(\hm\sigma)
+ \cos(l\sigma) \Pi S_0 \sinh(\hm\sigma) \right) \right. \\
&& \left.
+ {\hm\over\hom_l} \alpha^I \gamma^I
\left( \sin(l\sigma) \Pi S_0 \cosh(\hm\sigma)
+ \sin(l\sigma) S_0 \sinh(\hm\sigma) \right) \right]=0 \,. \nn
\label{zerodanger1}
\ea
Thus, the only terms involving any zero modes are
\ba
& 2 & {\hm\over\sqrt{X^+}} \int_0^\pi d\sigma  \left[
\left( y_1^I \sinh(\hm\sigma)
+  {y_2^I -y_1^I \cosh(\hm\pi)\over \sinh(\hm\pi)} \cosh(\hm\sigma)
\right) \right. \nn\\
& & \hspace{6cm} \gamma^I
\left( S_0' \cosh(\hm\sigma) + \Pi S_0' \sinh(\hm\sigma) \right)
\nn\\
&& \left.
+ \left( y_1^I \cosh(\hm\sigma)
+  {y_2^I -y_1^I \cosh(\hm\pi)\over \sinh(\hm\pi)} \sinh(\hm\sigma)
\right) \gamma^I
\left( \Pi S_0' \cosh(\hm\sigma) + S_0' \sinh(\hm\sigma) \right)
\right] \,.\nn
\ea
Performing  the $\sigma$-integrals gives
\be
{\cal Q}^0  =
{2\over\sqrt{X^+}}\, \left[ y_2^I \gamma^I
\left(\cosh(\pi\hm) S_0 + \sinh(\pi\hm) \Pi S_0 \right)
- y_1^I \gamma^I S_0 \right]\,.
\ee
The non-zero mode contribution proportional to
$c_n\,\alpha_l^I\,\gamma^I\, S_n e^{-i(\hom_l + \hom_n)\tau}$
with $n\ne \pm l$ arises with coefficient
\ba
{1\over 2\sqrt{X^+}} \int_0^\pi  d\sigma
&& \left\{\cos (l \sigma) \sin(n\sigma)
{4l\over \hom_l} \left( {\hm \over \hom_n}\,\Pi +
i {(\hom_n -n)\over \hom_n}\, \bbbone\right)\right. \nn\\
&& + \left.\sin (l \sigma)\cos(n\sigma)
{ 4\hm\over \hom_l} \left( {n \over \hom_n}\,\Pi +
i{n\, (\hom_n-n) \over \hom_n\, \hm}\,\bbbone\right)\right\}\, .
\label{dangterm}
\ea
Using the identities
\ba
\int_0^\pi \cos (l\sigma) \sin (n \sigma) &=& -{n\over n^2-l^2} \,
\left( (-1)^{n+l} -1\right)\nn\\
\int_0^\pi \sin (l\sigma) \cos (n \sigma) &=&   {l\over n^2-l^2} \left(
(-1)^{n+l}
-1\right)\,,
\label{trigid}
\ea
we see that each term in the sums in (\ref{dangterm})
vanishes. Similarly one can show that the terms with $n=l\ne 0$
vanish, and thus only terms with $n=-l$ contribute. This proves that
the supercharge is time independent. It is also easy to determine the
terms with $n=-l$ explicitly.  The resulting formula for the
supercharge is given in (\ref{superex}).

\section{Modular properties of the cylinder diagrams}
\label{modprops}

In this appendix the relation (\ref{gtrans1}) will be derived. In
the process of doing so, we shall also find an explicit expression for
$\widehat{\Delta}_{\hm}$.

\subsection{Closed-string perspective}
\label{closedpers}

The procedure for establishing the modular properties of $g_2^{(m)}$
begins by considering the logarithm of the expression (\ref{g2def})
and  performing a Poisson resummation over the integer $n$.  In order
to do this, it is important to rewrite (\ref{g2def}) so that the $n=0$
factor has the same form as the $n>0$ factors.  For this purpose it is
convenient to introduce a parameter $m_1$ and use the relation
\be
\lim_{m_1\to m}  \left(1-e^{-2\pi {\sqrt{m^2-m_1^2}}}\right)\,
\left( 1 + \left(m+m_1 \over m-m_1\right)
q^m \right)^{1\over 2} \,
\left( 1 + \left(m-m_1 \over m+m_1\right)
q^m \right)^{1\over 2}
= 4\pi\, m\, q^{m/2}\,.
\label{limdefs}
\ee
We can then write
\be
\ln  g_2^{(m)} = \lim_{m_1\to m}
\left(\calB_{m_1}   - \sum_{p=1}^{\infty}
{1\over p} e^{-2\pi p {\sqrt{m^2-m_1^2}}}  \right)
+2(2\pi t) \Delta_m   \, .
\label{logg2}
\ee
The term $\calB_{m_1}$ in (\ref{logg2}) is given by
\be
\calB_{m_1}  = - {1\over 2} \sum_{p=1}^{\infty}\sum_{n=-\infty}^\infty
 {(-1)^p\over p}  q^{\omega_n p}
\left[\left({\omega_n + m_1 \over \omega_n -m_1}\right)^p
+ \left({\omega_n - m_1 \over \omega_n +m_1}\right)^p \right]
\label{modefacs}
\ee
with $\omega_n = +\sqrt{m^2 + n^2}$. Next use the power series
expansion in powers of $\omega_n^2/m_1^2$,
\be
\left({\omega_n + m_1 \over \omega_n -m_1}\right)^p
+ \left({\omega_n - m_1 \over \omega_n +m_1}\right)^p
= \sum_{r=0}^{\infty} c_r^p
\left({\omega_n\over m_1}\right)^{2r}\,,
\label{powerser}
\ee
which converges for sufficiently large $m_1$ for any given value of
$n$. The coefficients $c_r^p$, which will not be
needed explicitly, can be used for smaller values of $m_1$ since the function
on the left-hand side is meromorphic. Using the identity  $e^{-z}
=\int_0^\infty ds\, e^{-s - \pi^2 z^2/4s}/\sqrt{\pi s}$,
(\ref{modefacs}) becomes
\ba
\calB_{m_1} && = - {1\over 2 \sqrt{\pi}} \sum_{n=-\infty}^\infty
\sum_{p=1}^{\infty} (-1)^p \sum_{r=0}^{\infty}
c_r^p \left({\omega_n\over m_1}\right)^{2r}
\int_0^\infty {ds\over s^{1/2}}
e^{-p^2 s -\pi^2t^2 \omega_n^2/s} \label{relogd} \\
&&=  - {1\over 2 \sqrt{\pi}} \sum_{n=-\infty}^\infty
\sum_{p=1}^{\infty} (-1)^p \sum_{r=0}^{\infty}
c_r^p {\partial^r \over (\partial  t^2)^r}
\int_0^\infty {ds\over s^{1/2}}
\left({-s\over \pi^2 m_1^2}\right)^r
e^{-p^2 s -\pi^2t^2 (m^2 + n^2)/s}\,.\nn
\ea
It is now easy to re-express the sum over $n$ by a Poisson
resummation which leads to a sum over $\hat n$,
\be
 \calB_{m_1} =  \calB'_{m_1} - {1\over 2 \pi}
\sum_{p=1}^{\infty} (-1)^p \sum_{r=0}^{\infty}
c_r^p {\partial^r \over (\partial  t^2)^r}
{1\over t} \int_0^\infty ds
\left({-s\over \pi^2 m_1^2}\right)^r
e^{-p^2 s -\pi^2t^2 m^2 /s} \, ,
\label{poissonres}
\ee
where the $\hat n > 0$ terms are contained in
\be
\calB'_{m_1} = - {1\over \pi} \sum_{\hn=1}^{\infty}
\sum_{p=1}^{\infty} (-1)^p \sum_{r=0}^{\infty}
c_r^p {\partial^r \over (\partial  t^2)^r}
{1\over t} \int_0^\infty ds
\left({-s\over \pi^2 m_1^2}\right)^r
e^{-p^2 s -\pi^2t^2 m^2 /s}  e^{-s\hn^2/t^2}\,  .
\label{bprimedef}
\ee

The $\hat n=0$ term is the second term in (\ref{poissonres})
which can be rewritten as
\be
- {\ttt \hm \over 2\pi} \sum_{p=1}^{\infty} (-1)^p \sum_{r=0}^{\infty}
c_r^p \left( {m \over m_1}\right)^{2r}
{\partial^r \over (\partial  \hm^2)^r}
{1\over \hm} \int_0^\infty ds
\left({-s\over \pi^2 }\right)^r
e^{-p^2 s -\pi^2 \hm^2 /s} \,.
\label{poissaga}
\ee
In the limit $m_1\rightarrow m$ this has the form
$2\pi \ttt\,  \widehat\Delta_\hm$, where $\widehat\Delta_\hm$ will be
identified with the off-set for the open string
\be
\widehat\Delta_\hm = - {1\over (2\pi)^2}
\sum_{p=1}^{\infty} (-1)^p \sum_{r=0}^{\infty}
c_r^p\, \hm\, {\partial^r \over (\partial  \hm^2)^r}
{1\over \hm} \int_0^\infty ds
\left({-s\over \pi^2 }\right)^r
e^{-p^2 s -\pi^2 \hm^2 /s} \,.
\label{offsett}
\ee
We can check that this expression reduces to the correct offset
for the flat space theory in the limit $m\to 0$.  In that limit
the only terms that survive are those in which $\partial^r / (\partial
\hm^2)^r$ acts on the $1/\hm$ factor, resulting in
\be
\widehat\Delta_0 =
\lim_{\hm \to 0} \sum_{p=1}^\infty -(-1)^p {-1\over 4\pi^2 p^2}
\sum_{r=0}^\infty c_r^p
\Gamma(2r+1) (-\hm^2\pi^2 p^2)^{-r}\, ,
\label{mzerd}
\ee
which can be written as
\ba
\widehat\Delta_0 && =
\lim_{\hm\to 0}  \sum_{p=1}^\infty (-1)^p {-1\over 4\pi^2 p^2}
 \int_0^\infty dy \,e^{-y}\, y^{2r}
\sum_{r=0}^\infty c_r^p (-\hm^2 \pi^2 p^2)^{-r} \nn\\
 &&= \lim_{\hm\to 0}
\sum_{p=1}^\infty (-1)^p {-1\over 4\pi^2 p^2} \int_0^\infty dy \,
e^{-y}\,
\left[\left({y^2 - \pi^2 p^2 \hm^2\over  y^2 + \pi^2 p^2 \hm^2}
\right)^p
+ \left({y^2 + \pi^2 p^2 \hm^2\over  y^2- \pi^2 p^2 \hm^2} \right)^p
\right] \nn\\
  &&= 2\sum_{p=1}^\infty (-1)^p {-1\over 4\pi^2 p^2} = {1\over 24} \, ,
\label{newwrite}
\ea
where the fact that $\sum_{p=1}^\infty (-1)^p/p^2 = - \pi^2 /12$
has been used in the last step.  This agrees with the standard
expression for the contribution from the Casimir energy to the function
$f_4$ in the flat space case.

The $\hat n>0$ terms in (\ref{poissonres}) can be evaluated explicitly
to give (defining $\ts=s/t^2$),
\ba
\calB'_{m_1} = &&
-{1\over \pi} \sum_{\hn=1}^{\infty}
\sum_{p=1}^{\infty} \sum_{r=0}^{\infty}(-1)^p c_r^p  \int_0^\infty d\ts
\sum_{l=0}^{\infty} {(-p^2 \ts )^l \over l!} \nn\\
& & \qquad
{\Gamma(l+r+{3\over 2}) \over \Gamma(l+{3\over 2})}
\left({- \ts\over \pi^2 m_1^2}\right)^r
 t^{2l+1}
e^{-\hn^2 \ts -\pi^2 m^2 /\ts}\,.
\label{firstlin}
\ea

{}From the definition (\ref{logg2}) it is clear that both
$\calB_{m_1}$ and the term
$\sum_{p=1}^{\infty} e^{-2\pi p{\sqrt{m^2-m_1^2}}}/p$ in (\ref{logg2})
are separately divergent in the limit  $m_1\to m$. For this reason it
is important to keep $m_1 \ne m$ in (\ref{firstlin}). We have
now reexpressed $\calB_{m_1}$ in (\ref{logg2}) in a form that will be
compared with the expression obtained from the open string
calculation.

\subsection{Open-string perspective}
\label{openpers}

We now start with the expression $\hg_4^{(m)}$ in (\ref{g4hatdeff}).
In this case the logarithm has the form
\be
\ln \hg^{(\hm)}_4 =   \calC + 2\pi\ttt\, \widehat{\Delta}_\hm   \,,
\label{logg4}
\ee
where
\be
\calC_{m_1} =
- {1\over 2 \sqrt{\pi}} \sum_{\hn=1}^{\infty}
 \sum_{\hp} \int_0^\infty {d\ts\over \ts^{1/2}}
e^{-\hn^2 \ts} e^{-\pi^2 \ttt^2 (\hm^2+\hp^2)/\ts} \,.
\label{logzo}
\ee
Here the sum over $\hp$ runs over both ${\cal P}_+$ and
${\cal P}_-$, including the value  $\hat p=0$ in both sectors. The
second term on the right-hand side of
(\ref{logg4}) should be given by the Casimir energy
of a two-dimensional boson field on an open-string world-sheet of
width $\ttt$ with the appropriate boundary conditions.
We will postpone the discussion  of this term until later when
we will see that it coincides  with the expression obtained from the
closed-string side in the second term in (\ref{poissonres}).

We will start by considering $\calC_{m_1}$. In order to proceed we
need to perform a Poisson resummation over the values of $\hp$. This
is achieved by replacing each term in the $\hp$ sum by a contour
integral over a complex variable $\rho$, enclosing the relevant pole
at the value of $\hp$ that solves the above relations, giving
\ba
\calC_{m_1} = && - {1\over 2 \pi \sqrt{\pi}}
\oint d\rho \sum_{\hn=1}^{\infty}
\int_0^\infty {d\ts \over \ts^{1/2}}
e^{-\hn^2 \ts - \pi^2 \ttt^2 (\hm^2+\rho^2)/\ts}\nn\\
&& \qquad\qquad \left[
{  \left( \pi + {\hm_1 \over \rho^2 + \hm_1^2} \right) \over
1 + \left({\rho+i\hm_1 \over \rho-i\hm_1}\right) e^{-2\pi i \rho}} +
{ \left( \pi - {\hm_1 \over \rho^2 + \hm_1^2} \right) \over
1 + \left({\rho-i\hm_1 \over \rho+i\hm_1}\right) e^{-2\pi i \rho}} \right]
\,.
\label{logzoo}
\ea
The contour is the sum of  small circles enclosing all the poles
arising from real zeroes of the denominators in the square brackets on
the right-hand side of this equation.  The initial contour may now be
deformed into the sum of two straight lines. One of these is a
straight line, $L_1$,  displaced by $\epsilon>0$ above
the real axis and running from $\rho = \infty + i\epsilon$ to
$\rho = -\infty + i\epsilon$.  The other part is the line $L_2$,
displaced by $-\epsilon$ below the real axis and running from
$\rho = -\infty - i\epsilon$ to $\rho =  \infty - i\epsilon$.

In order to expand the denominators of the terms in the square
brackets in (\ref{logzoo}) in convergent series it is important
to choose  $\epsilon > \hm_1$. This means that the upper and lower
contours are shifted into the upper and lower half $\rho$-plane,
respectively.  It is therefore
important to understand the singularity structure
of the integrand in the complex $\rho$-plane in (\ref{logzoo}).
A careful analysis reveals that the
denominators of the terms in square brackets
have no zeroes or singularities away from the real $\rho$ axis.
However, the numerators of the
terms in the square brackets have poles at $\rho = \pm i\hm_1$
and the residues of these poles contribute when the integration
contours are displaced.  The total contribution from these two poles
to $\calC_{m_1}$ is
\be
\calC_{m_1}^{\prime\prime} = - {1\over \sqrt \pi}
\int_0^\infty {d\ts \over \ts^{1/2}}
e^{-\hn^2 \ts - \pi^2 \ttt^2 (\hm^2 - \hm_1^2)/\ts}
=  - \sum_{\hn = 1}^\infty {1\over \hn}e^{-2\pi \hn \sqrt{m^2 -m_1^2}}
\,,
\label{polecons}
\ee
which matches the second term in brackets in (\ref{logg2}).
The total value of $\calC_{m_1}$ is therefore
\be
\calC_{m_1} = \calC_{m_1}^{L_1} + \calC_{m_1}^{L_2} +
\calC_{m_1}^{\prime\prime} \,,
\label{twocons}
\ee
where $\calC_{m_1}^{L_1}$ and $\calC_{m_1}^{L_2}$ denote the terms
coming from the integration over the upper and lower contours.

Along the lower contour $L_2$ we expand the denominators in the
square brackets of (\ref{logzoo}) in powers of
$({\rho +i\hm_1 /\rho-i\hm_1}) e^{-2\pi i \rho}$, which gives
\ba
\calC_{m_1}^{L_2} = && -{1\over 2 \pi \sqrt{\pi}}
\int_{-\infty}^{\infty} d\rho \sum_{\hn=1}^{\infty}
\int_0^\infty {d\ts \over \ts^{1/2}}
\sum_{p=0}^{\infty} (-1)^p
e^{-\hn^2 \ts - \pi^2 \ttt^2 (\hm^2+\rho^2)/\ts} e^{-2\pi i \rho p}\nn \\
&& \qquad
\left[ \left( \pi + {\hm_1 \over \rho^2 + \hm_1^2} \right)
\left({\rho+i\hm_1 \over \rho-i\hm_1}\right)^p +
\left( \pi - {\hm_1 \over \rho^2 + \hm_1^2} \right)
\left({\rho-i\hm_1 \over \rho+i\hm_1}\right)^p \right] \,.
\label{lowercont}
\ea
In order to obtain a convergent series expansion along the upper
contour $L_1$ it is necessary to expand in powers of
$({\rho -i\hm_1 /\rho+i\hm_1}) e^{2\pi i \rho}$.  After taking into
account the reversal of orientation of the integration this gives
$\calC_{m_1}^{L_1}$ of the same form as $\calC_{m_1}^{L_2}$,  except
that now the
sum over $p$ does not include the $p=0$ term,
\be
\calC_{m_1}^{L_2} - \calC_{m_1}^{L_1} = - {1\over \sqrt{\pi}}
\int_{-\infty}^{\infty} d\rho \sum_{\hn=1}^{\infty}
\int_0^\infty {d\ts \over \ts^{1/2}}
e^{-\hn^2 \ts - \pi^2 \ttt^2 (\hm^2+\rho^2)/\ts} \,.
\label{uplow}
\ee

We will separate out the  $p=0$ term by writing
\be
\calC_{m_1} \equiv  \calC_{m_1}^{L_1}+
\calC_{m_1}^{L_2}+ \calC_{m_1}^{\prime\prime} =\calC'_{m_1}
+  \calC_{m_1}^{\prime\prime} + 2 (2\pi t) \Delta_m ,
\label{sepcs}
\ee
where
\ba
2 (2\pi t) \Delta_m  = &&
- {1\over \sqrt{\pi}}
\int_{-\infty}^{\infty} d\rho \sum_{\hn=1}^{\infty}
\int_0^\infty {d\ts \over \ts^{1/2}}
e^{-\hn^2 \ts - \pi^2 \ttt^2 (\hm^2+\rho^2)/\ts}\nn\\
 = &&
- {1\over \ttt \pi} \sum_{\hn=1}^{\infty} \int_0^\infty d\ts\,
e^{-\hn^2 \ts - \pi^2 \ttt^2 \hm^2 / \ts}\nn\\
 = &&  - 2 (2\pi t) {1\over (2\pi)^2}
\sum_{\hn=1}^{\infty} \int_0^\infty d\ts\,
e^{-\hn^2 \ts - \pi^2 m^2 / \ts} \, .
\label{poterm}
\ea
thus producing the correct off-set of the closed string calculation.
Next expand the $p\ne 0$ terms in a power series in $\rho/\hm_1$. The
terms with coefficient $\pi$ in the first bracket in (\ref{lowercont})
only involve powers of $\rho^2/\hm_1^2$.  For these terms we have the
expansion
\be
2 \cos 2p \phi \equiv
\left({\rho + i\hm_1 \over \rho-i\hm_1}\right)^p  +
\left({\rho - i\hm_1 \over \rho+i\hm_1}\right)^p
= \sum_r c_r^p (-1)^r (\rho^2/\hm_1^2)^r\,,
\label{cosdef}
\ee
where the coefficients are the {\it same} as those that arose in the
closed string calculation.
Note that we can also write
\be
 2i \sin 2p \phi \equiv
\left({\rho + i\hm_1 \over \rho-i\hm_1}\right)^p
- \left({\rho - i\hm_1 \over \rho+i\hm_1}\right)^p
= - {i\over p} {d \cos(2p\phi) \over d \phi} =
- {i\over p} {d \cos (2p \phi) \over d\rho}
\left({d\phi \over d \rho}\right)^{-1} \,.
\label{sindef}
\ee
Using
\be
e^{2i\phi} =\left({\rho + i\hm_1 \over \rho-i\hm_1}\right)
\label{identphi}
\ee
we see that
\be
\left({d\phi \over d \rho}\right) =
- {\hm_1\over \rho^2 + \hm_1^2}\,.
\label{diffedef}
\ee
Substituting these relations into the expression for $\calC_{m_1}$
gives
\be
\calC_{m_1}' = \calC_{m_1}^e + \calC_{m_1}^o\,,
\label{lnzo}
\ee
where
\be
\calC_{m_1}^e = - {1 \over  \pi \sqrt{\pi}}
\int_{-\infty}^{\infty} d\rho \sum_{\hn=1}^{\infty}
\int_0^\infty {d\ts \over \ts^{1/2}}
\sum_{p=1}^{\infty} (-1)^p
e^{-\hn^2 \ts - \pi^2 \ttt^2 (\hm^2+\rho^2)/\ts} e^{-2\pi i \rho p}
2 \pi \cos(2p\phi) \,,
\label{lnzoedef}
\ee
and
\be
\calC_{m_1}^o = - {1 \over  \pi \sqrt{\pi}}
\int_{-\infty}^{\infty} d\rho \sum_{\hn=1}^{\infty}
\int_0^\infty {d\ts \over \ts^{1/2}}
\sum_{p=1}^{\infty} (-1)^p
e^{-\hn^2 \ts - \pi^2 \ttt^2 (\hm^2+\rho^2)/\ts} e^{-2\pi i \rho p}
{i\over p} {d \cos (2p \phi) \over d\rho}
 \,.
\label{lnzoodef}
\ee
Integrating the last integral by parts gives two terms where the
$\partial/\partial \rho$ acts on $e^{-2\pi i \rho p}$ and on
$e^{-\pi^2\ttt^2\rho^2/\ts}$.
Therefore the full expression for $\calC_{m_1}'$ is
\ba
\calC_{m_1}'  && = {i\over  \pi \sqrt{\pi}}
\int_{-\infty}^\infty d\rho \sum_{\hn=1}^\infty
\int_0^\infty {d\ts \over \ts^{1/2}}
\sum_{p=1}^{\infty}  {(-1)^p\over p}
e^{-2\pi i \rho p} \, \cos 2p \phi \,
{d  \over d\rho}
e^{-{\pi^2\ttt^2\over \ts}(\rho^2 +\hm^2) -\hn^2\ts}\label{fullzo}\\
&& = - 2 i \sqrt{\pi}
\int_{-\infty}^\infty d\rho\, \rho\, \sum_{\hn=1}^\infty
\int_0^\infty {d\ts \over \ts^{3/2}}
\sum_{p=1}^{\infty}  {(-1)^p \over p}\, \ttt^2\,
e^{-2\pi i \rho p}\, \cos 2p \phi \,
e^{-{\pi^2\ttt^2\over \ts}(\rho^2 +\hm^2) -\hn^2\ts}\,.
\nn
\ea
Substituting the expansion for $\cos 2p\phi$ gives
\ba
\calC_{m_1}'  &=& - i \sqrt{\pi}
\int_{-\infty}^\infty d\rho\,  \sum_{\hn=1}^\infty
\int_0^\infty {d\ts \over \ts^{3/2}}
\sum_{p=1}^{\infty}  {(-1)^p\over p}\, \ttt^2\nn\\
&& \qquad
\sum_{r=0}^{\infty} c_r^p (-1)^r \left({\rho^2\over\hm_1^2}\right)^r
\, \rho\, e^{-2\pi i \rho p}
e^{-{\pi^2\ttt^2\over \ts}(\rho^2 +\hm^2) -\hn^2\ts}\,.
\label{zofin}
\ea
We can replace the odd power of $\rho$ with the factor
$i\partial/ 2\pi \partial p$ acting only on the exponential to its
right.  The remaining powers of $\rho^2$ can be replaced by
derivatives with respect to $\ttt^2$.   This gives
\ba
\calC_{m_1}'  = && {1\over 2 \sqrt{\pi}}
\int_{-\infty}^\infty d\rho  \sum_{\hn=1}^\infty
\int_0^\infty {d\ts \over \ts^{3/2}}
\sum_{p=1}^{\infty}  {(-1)^p\over p} \ttt^2\nn\\
&& \qquad \sum_{r=0}^{\infty} c_r^p
\left({\ts \over \pi^2\hm_1^2}\right)^r\,
e^{-{\pi^2 \ttt^2 \hm^2 \over \ts}}
{\partial^r \over (\partial \ttt^2)^r}\,  {\partial\over \partial p}\,
e^{-2\pi i \rho p} \,
e^{-{\pi^2\ttt^2\rho^2\over \ts}  -\hn^2\ts}\,.
\label{newlogzz}
\ea
We can now do the $\rho$ integral very simply by completing a square,
changing variables from $\rho$ to
 $\rho' = \rho - i {p \tilde s/\pi \ttt^2}$ and shifting the
 $\rho'$ integration contour to lie along the real axis,
\ba
\calC_{m_1}'  && = {1\over 2 \pi}
\sum_{\hn=1}^\infty
\int_0^\infty {d\ts \over \ts}
\sum_{p=1}^{\infty}  {(-1)^p\over p}\, \ttt^2
\sum_{r=0}^{\infty} c_r^p
\left({\ts \over \pi^2\hm_1^2}\right)^r\,
e^{-{\pi^2 \ttt^2 \hm^2 \over \ts}}
{\partial^r \over (\partial \ttt^2)^r}\,  {\partial\over \partial p}\,
{1\over \ttt}
e^{-{p^2 \ts\over \ttt^2}  -\hn^2\ts}\nn\\
&& = - {1\over \pi}
\sum_{\hn=1}^\infty
\int_0^\infty d\ts \,
\sum_{p=1}^{\infty}  (-1)^p \, \ttt^2
\sum_{r=0}^{\infty} c_r^p
\left({ \ts \ttt^2 \over \pi^2 m_1^2}\right)^r\,
e^{-\pi^2 m^2 / \ts}
{\partial^r \over (\partial \ttt^2)^r}\,
{1\over \ttt^3}
e^{-{p^2 \ts\over \ttt^2}  -\hn^2\ts} ,
\label{squarecom}
\ea
where $\hm_1=m_1/\ttt$ has been used in the final line.
Expanding the exponential involving $\ttt$, and differentiating term
by term, gives
\ba
\calC_{m_1}' =&&
- {1\over \pi}
\sum_{\hn=1}^\infty
\int_0^\infty d\ts \,
\sum_{p=1}^{\infty}  (-1)^p \,
\sum_{r=0}^{\infty} c_r^p
\left({  \ts \over \pi^2 m_1^2}\right)^r\,
e^{-\pi^2 m^2 / \ts - \hn^2 \ts} \nn\\
&& \qquad \qquad
\sum_{l=0}^{\infty} {(-p^2 \ts)^l \over l!}\, {\Gamma(-l-1/2)\over
\Gamma(-l-r-1/2)}
 \, \ttt^{-(2l+1)}\nn\\
=&& - {1\over \pi}
\sum_{\hn=1}^\infty
\int_0^\infty d\ts \,
\sum_{p=1}^{\infty}  (-1)^p \,
\sum_{r=0}^{\infty} c_r^p
\left({ - \ts \over \pi^2 m_1^2}\right)^r\,
e^{-\pi^2 m^2 / \ts - \hn^2 \ts} \nn \\
&& \qquad \qquad
\sum_{l=0}^{\infty} {(-p^2 \ts)^l \over l!}\, {\Gamma(l+r+1/2)\over
\Gamma(l+ 3/2)}
 \, \ttt^{-(2l+1)}\, .\nn\\
\label{difftt}
\ea
This reproduces (\ref{firstlin}) upon setting $\ttt=1/t$.

We now turn to the second term in (\ref{logg4}), which is
proportional to the Casimir energy of a two-dimensional field.
This is equal to the difference of the vacuum energy on a finite
strip of width $\ttt$ and an infinitely wide strip.  Using a
standard argument, the vacuum energy on the finite width strip is
given by setting $\hn =0$  in (\ref{logzo}).  Each term in the
resulting  series in the $\hp$ sum is divergent.  However, after the
Poisson resummation the divergence is entirely in the $p=0$ term,
which, in the $m_1\to m$ limit, is the $\ttt$-independent term
that has to be subtracted in obtaining the
Casimir energy.  Therefore,  the Casimir energy is obtained by
setting $\hn=0$ in the summand of the right-hand side of
(\ref{squarecom}), which gives
\be
2\pi \ttt \widehat \Delta =  - {1\over \pi}
\int_0^\infty d\ts \,
\sum_{p=1}^{\infty}  (-1)^p \, \ttt^2
\sum_{r=0}^{\infty} c_r^p
\left({ \ts \ttt^2 \over \pi^2 m_1^2}\right)^r\,
e^{-\pi^2 m^2 / \ts}
{\partial^r \over (\partial \ttt^2)^r}\,
{1\over \ttt^3}
e^{-{p^2 \ts\over \ttt^2}}\, ,
\label{openoffsett}
\ee
thus reproducing (\ref{offsett}). This completes the argument.

\end{document}